\newcommand{\rem}[1]{}
\newcommand{\eqref}[1]{(\ref{#1})}
\documentclass[12pt]{iopart}
\usepackage{amsfonts}
\usepackage{hyperref}
\usepackage{epsfig}

\usepackage{subfigure}
\usepackage{scalebar}

\begin{document}

\title{Carbon $sp$ chains in graphene nanoholes}

\author{Ivano E.\ Castelli$^{1,2}$, Nicola Ferri$^{2}$,
Giovanni Onida$^{2}$, and Nicola Manini$^{2}$}
\address{$^1$
Center for Atomic-scale Materials Design,
Department of Physics, Technical University of Denmark,
DK-2800 Kongens Lyngby, Denmark}
\address{$^2$
ETSF and Dipartimento di Fisica, Universit\`a di Milano,
Via Celoria 16, 20133 Milano, Italy}

\date{\today}

\begin{abstract}
Nowadays $sp$ carbon chains terminated by graphene or
graphitic-like carbon are synthesized routinely in several nanotech labs.
We propose an {\it ab-initio} study of such carbon-only materials, by
computing their structure and stability, as well as their electronic,
vibrational and magnetic properties.
We adopt a fair compromise of microscopic realism with a certain level of
idealization in the model configurations, and predict a number of
properties susceptible to comparison with experiment.
\end{abstract}


\section{Introduction}
\label{intro:sec}

The carbon atom, with its three possible hybridization states, originates
in nature very different elemental materials.
The three possibilities ($sp^3$, $sp^2$, $sp$) correspond to three
different prototypical structures: respectively diamond, graphitic-like
structures (such as graphite, graphene, carbon nanotubes, and fullerenes),
and linear carbon chains (known in the literature as {\it polyynes}
\cite{Cataldo05,polyynes:note} or $sp$ carbon chains, spCCs in short).

spCCs were discovered in nature around 1968 \cite{ElGoresy68}, but their
role in the arena of carbon-based nanostructures has been quite marginal:
till very recently, spCCs have been considered as exotic allotropic
forms, mainly present in extraterrestrial environments.
Indeed, in the interstellar clouds formed in the explosions of carbon
stars, novae, and supernovae, spCCs have been detected aside with
fullerenes \cite{Kroto92,Duley09}, amorphous carbon dust, cyanopolyynes,
and oligopolyynes.
Indeed, in the phase diagram of carbon, the field of existence of spCCs
coincides with that of fullerenes~\cite{Bundy96}.

The high reactivity of spCCs \cite{Baughman06}, and their tendency to
undergo cross-linking to form sp$^2$ structures, directed the experimental
efforts towards complicate strategies for the stabilization of the spCCs
with molecular end-groups or their isolation in inert matrices
\cite{Cataldo05,Matsuda84,Kudryavtsev69}.
Thanks to novel synthetic routes and strategies \cite{Cataldo99}, polyynes
of increasing length and different type of termination have been
successfully synthesized and characterized
\cite{Mohr03,Zhao03,Liu03,Inoue10,Rice10}.
End-capped spCCs (often, but not exclusively, hydrogen-terminated) are
being synthesized by chemical / electrochemical / photochemical methods
\cite{Cataldo05, Kijima96, Heimann99, Tsuji03, Cataldo04}.
spCCs have also been produced from carbon by dynamic pressure
\cite{Heimann99,Yamada91}.
spCCs or carbynoid material containing up to $300$ carbon atoms were
synthesized, which opens a promising route toward molecular engineering of
$sp$-carbon structures \cite{Kijima96, Heimann99, Ohmura97,Kijima97}.

Samples of pure carbon films grown by supersonic cluster beam deposition at
room temperature have been characterized, and proven to contain a sizeable
$sp$ component \cite{Ravagnan02,Ravagnan07}.
These spCCs showed weaker stability relative to the $sp^2$ component upon
exposition to low pressure gases at room temperature \cite{Ravagnan09}.
spCCs have also been produced from stretched nanotubes \cite{Troiani03}.
Recently, Jin {\it et al.}~\cite{Jin09} have realized spCCs by stretching
and thinning a graphene nanoribbon from its two free edges, by removing
carbon rows until the number of rows becomes one or two.
These spCCs show a good stability under the beam of a transmission electron
microscope (TEM) for lengths up to a few nanometers.
These experimental observations of spCCs formation during the controlled
electron irradiation of graphene planes resulted in an rapidly increasing
interest in this field
\cite{Chuvilin09,Mikhailovskij09,Zeng10,Chalifoux09,Hobi10,Akdim11,Hu11,Ravagnan11,Erdogan11}.

Meanwhile, the existence of intrinsic magnetism in pure carbon has been a
matter of debate for quite some time \cite{Makarova01}.
Possible effects due to magnetic contaminants on the experimental results
have been discussed \cite{Esquinazi02}.
Subsequently, it has been shown that possible contamination effects are
unable to explain quantitatively the measured ferromagnetism, supporting
the idea that carbon magnetism has an intrinsic origin \cite{Coey02}. 
The existence of $\pi$-electron magnetism in pure carbon has now been
widely accepted (see e.g. \cite{Esquinazi03, Ohldag07}). 
On the theoretical side, it is well known that magnetic instabilities exist
at specific graphene edges \cite{Klein99,Son06b,Yazyev08,Uchoa08}, in
defective graphene \cite{Pisani08} and  nanotubes \cite{Zanolli10}.

The discovery of pure-carbon $\pi$-electron magnetism has also lead to
speculations about possible applications of carbon-based magnetic materials
in molecular electronic devices: for example, spintronic devices built
around the phenomenon of spin-polarization localized at the 1-D zig-zag
edges of graphene have been proposed \cite{Son06}.
Recently, the accent has been also put on the possibility to modify the
magnetization optically \cite{Yang08b}.
Indeed, linear carbon chains are nowadays considered promising structures
for nano-electronic applications \cite{Standley08,YLi08}.
For example, they can be used as molecular bridges across graphene nanogap
devices.
Potential applications for the realization of non-volatile memories and
two-terminal atomic-scale switches have been demonstrated \cite{YLi08}.

The remarkable robustness of spCCs terminated on pure carbon, graphene-like
fragments, combined with the fast progresses in the synthesis of graphene
and graphene derivatives, could open the way towards the realization of
actual nanodevices based on $sp$ + $sp^2$ carbon nanostructures.
The interest of such systems stems from the possibility to exploit their
peculiar semiconducting-magnetic behavior, in order to achieve novel
characteristics and functions for the target devices.

Clearly, the possibility of designing graphene-based magnetic
nanostructures is particularly intriguing.
The capability of arranging the spins inside a carbon structure in a
variety of ways, could open the way for the construction of completely
novel devices \cite{DasSarma01}.
Possible future applications for spCCs in interaction with the
graphene-type system could be the construction of microchips with
ferromagnetic or antiferromagnetic character that can be controlled by
nanomanipulation and read out by nanocurrents.

In Sect.~\ref{model:sec} we introduce the theoretical model, whose
electronic structure we address by a standard {\it ab-initio} method based
on the Density-Functional Theory (DFT).
Section \ref{carb_bind:sec} presents the investigation of the structural
and binding properties of a spCC inside a nanohole (nh) in a graphene sheet
representing the $sp^2$ component in a carbon-only sample.
Section \ref{mag_graph:sec} studies the magnetic properties of the nh edges
and of the inserted spCCs.
In Sections~\ref{elect_graph:sec} and \ref{vibr_graph:sec} we cover the band
structures and selected vibrational properties of the studied
nanostructures.
In Sect.~\ref{highT:sec} we investigate the dynamical stability of the
metastable spCC-nh structures, by means of a tight-binding molecular
dynamics model.
Section~\ref{conclusion:sec} discusses the results of simulations in the
light of experimental data.

\section{The model}\label{model:sec}

In the present paper, we focus on spCCs bound to graphitic structures,
represented by a hole in a infinite graphene sheet.  This system is
representative of a class of $sp + sp^2$ systems, which are at the core of
an intense experimental work \cite{Ravagnan02,Ravagnan07,Ravagnan09,
  Jin09,Chuvilin09,Casari04,Cataldo10, Cinquanta11}.
To address the structural, vibrational, and electronic properties of spCCs
inserted into a nanometer-sized hole defect of a graphene layer, we resort
to the DFT.
The plane-wave pseudopotential method and the Local Spin-Density
Approximation (LSDA) to DFT have provided a simple framework whose accuracy
have been demonstrated in a variety of systems \cite{Pickett89}.
The time-honored LDA is one in many functionals being used for current DFT
studies of molecular and solid-state systems: other functionals often
improve one or another of the systematic defects of LDA (underestimation of
the energy gap, small overbinding and tiny overestimation of the
vibrational frequencies), but to date no functional is universally accepted
to provide systematically better accuracy than LDA for all properties of
arbitrary systems.
For a covalent system of $s$ and $p$ electrons as the one studied here, LDA
is appropriate, and we expect our results to change by a few percent at
most if the calculations were repeated using some other popular
functional \cite{B3LYP,PBE96,Xu04}.

We compute the total adiabatic energies by means of the code Quantum
Espresso \cite{espresso2009}, which computes forces by standard
Hellmann-Feynman method.
Each self-consistent electronic-structure calculation stops when the total
energy changes by less than $10^{-8}$~Ry.
We use ultrasoft pseudopotentials \cite{Vanderbilt90,Favot99}, for which a
moderate cutoff for the wave function/charge density of $30/240$~Ry is
sufficient.
We terminate atomic relaxation when all residual force components are
smaller than $10^{-4}$~Ry$/a_0\simeq 0.04\,\mu$N.

\begin{figure}
\begin{center}
\includegraphics[width=0.65\textwidth,angle=0,clip=]{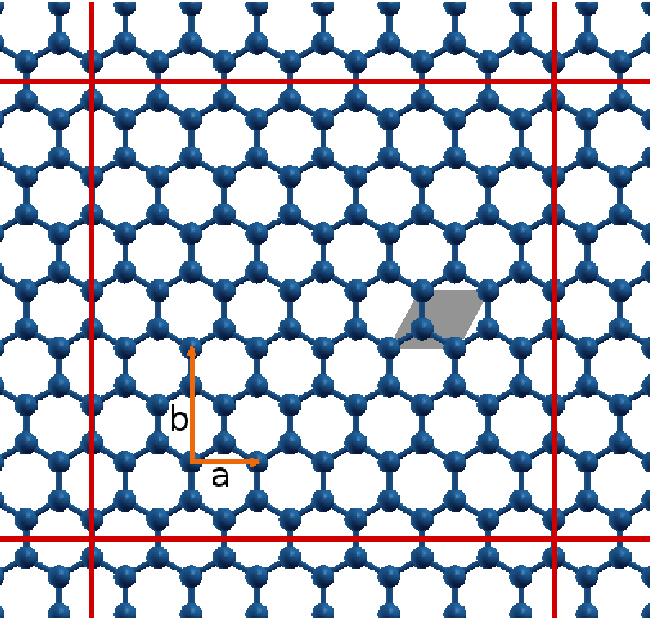}
\end{center}
\caption{\label{slab_pos:fig} (Color online)
  The adopted slab model supercell for graphene, composed by $7\times 4$
  conventional rectangular cells containing 4-atoms, each equivalent to two
  primitive unit cells of graphene (in gray).
  Solid: the unit vectors of the conventional cell, of length $a=244$~pm
  and $b=a \sqrt 3 \simeq 422$~pm.
  The size of the full supercell (delimited by the solid lines) in the
  $x-y$ plane is $1714$\,pm~$\times~1690$\,pm.
}
\end{figure}

Plane waves require periodic boundary conditions in all space directions.
In our model, we represent a graphene plane in a $(x-y)$-periodically
repeated $1714\times1690$~pm$^2$ supercell consisting of $7\times 4$
rectangular conventional unit cells containing four carbon atoms each, see
Fig.~\ref{slab_pos:fig}.
Within this cell, a complete graphene layer is then represented by $112$
carbon atoms.
We represent the graphite surface in the slab approximation, as one or a
few stacked graphene layers: to ensure that the interaction between
periodic images of the graphene sheet is negligible we interpose at least
$1$~nm of vacuum along the $z$ direction.
We have selected this cell size as it allows us to create reasonably large
nanoholes in the graphene layers, with fairly small interaction between the
supercell repeated copies of the nanohole itself, at the price of a
manageable computer time.
We cover electronic band dispersion the electron bands in the horizontal
plane by means of a $5\times 5\times 1$ $\Gamma$-centered {\bf k}-point
mesh.
BZ integration of the metallic band energies are performed using a $2
\times 10^{-4}$~Ry-wide Gaussian smearing of the fermionic occupations.

\subsection{The Nanohole}\label{nanohole:sec}

\begin{figure}
\begin{center}
  \subfigure[Initial]{\includegraphics[width=0.35\textwidth,angle=0,clip=]{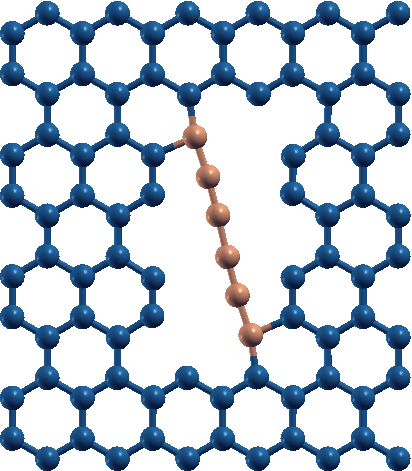}\label{little_in:fig}}
  \hspace{2.5mm}
  \subfigure[Relaxed]{\includegraphics[width=0.35\textwidth,angle=0,clip=]{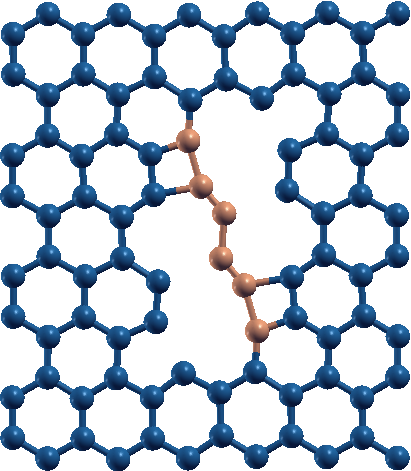}\label{little_out:fig}}
\end{center}
\caption{\label{hole_little:fig} (Color online)
  Initial and relaxed positions of a small nh with a C$_6$ polyyne: the
  lateral nh size is so small that no barrier prevents two polyyne atoms
  (brown/clear) to bind to the armchair edges.
}
\end{figure}

\begin{figure}
\begin{center}
  \subfigure[Initial]{\includegraphics[width=0.35\textwidth,angle=0,clip=]{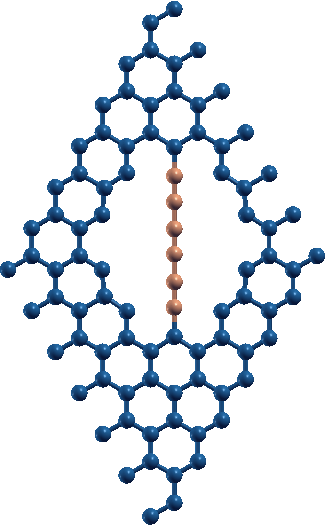}\label{esa_in:fig}}
  \hspace{2.5mm}
  \subfigure[Relaxed]{\includegraphics[width=0.35\textwidth,angle=0,clip=]{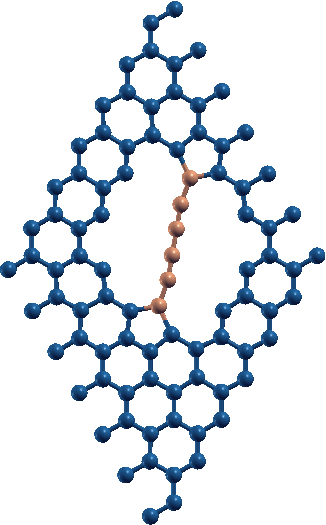}\label{esa_out:fig}}
\end{center}
\caption{\label{hole_esa:fig} (Color online)
  Initial and relaxed positions of a small nh with a C$_6$ polyyne: the
  lateral nh size is so small that no barrier prevents two polyyne atoms
  (brown/clear) to bind to the zig-zag edges and form pentagonal rings.
}
\end{figure}

Starting from the perfect graphene foil of Fig.~\ref{slab_pos:fig}, we
remove selected atoms in order to form a nh.
The size of the nh should be such that inserted spCCs fit and only bind at
their ends.
If the nh is too small, then spCC atoms would tend to reconstruct
the edges of the hole, as illustrated in the examples of
Figs.~\ref{hole_little:fig} and \ref{hole_esa:fig}.
There, the small spCC-edge distance leads to spontaneous (barrier-free)
edge reconstruction with the formation of additional squares or pentagons.

\begin{figure}
\centerline{
\includegraphics[width=0.5\textwidth,angle=0,clip=]{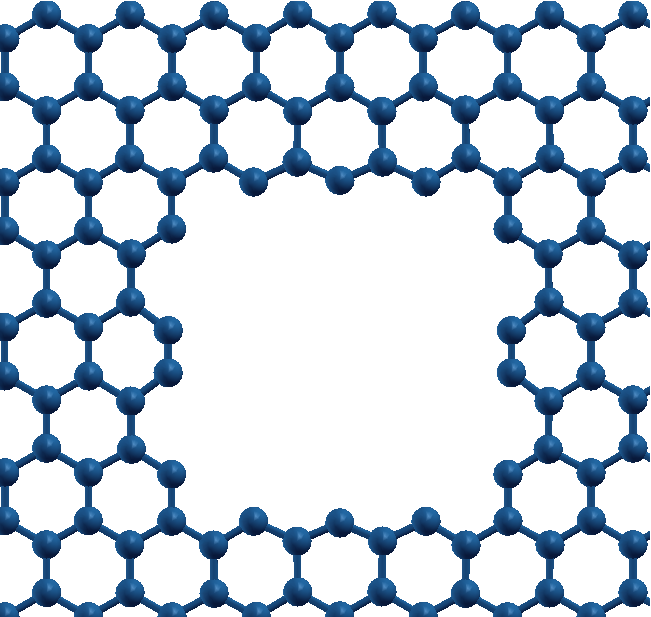}
}
\caption{\label{hole_pos:fig} (Color online)
  A nh of sufficient size to lodge spCCs is obtained by removing $28$
  atoms from the perfect graphene of Fig.~\ref{slab_pos:fig}.
  The rectangular hole has size $975$\,pm~$\times~985$\,pm, and is
  delimited by both zig-zag and armchair edges
}
\end{figure}

We must therefore construct a large enough nh, in particular allowing for a
distance of at least $300$~pm between the spCC and the nearest nh edge in
order to prevent recombination reactions.
Figure~\ref{hole_pos:fig} shows the minimal nh with such property.
Starting from the perfect graphene of Fig.~\ref{slab_pos:fig}, the nh is
obtained by removing $28$ atoms forming a rectangular hole of size
$975$\,pm~$\times~985$\,pm with both zig-zag armchair edges.
The size of the nh permits us to investigate the insertion of several
spCCs, from C$_5$ to C$_8$ atoms long, with various positions relative
to the nh edges.

\section{spCCs Binding to a Nanohole}\label{carb_bind:sec}

\begin{figure}
\centerline{
\includegraphics[width=0.5\textwidth,angle=0,clip=]{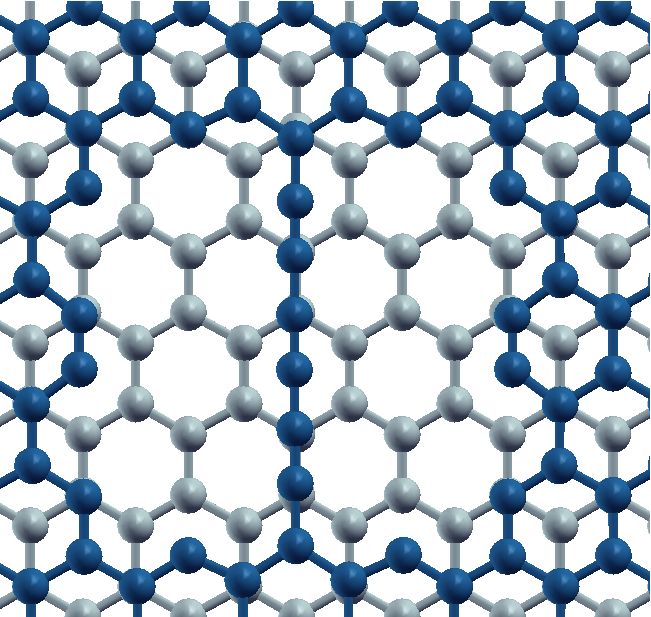}}
\caption{\label{carb6_2lay:fig} (Color online)
  A nano-indented graphite surface represented by a perfect graphene foil
  (pale gray) plus a superposed sheet with a nh (dark blue) as in
  Fig.~\ref{hole_pos:fig} stacked in the standard AB arrangement of
  graphite.
  A C$_6$ chain (also dark blue) is connected to the zig-zag edge of the nh.
  The layer-layer distance equals $332$~pm.
}
\end{figure} 

In an experiment \cite{Ravagnan02,Ravagnan07,Ravagnan09,Casari04}, spCCs
are likely to be bound to extended $sp^2$ structures, more reminiscent of
graphite than of graphene.
Such a configuration could be described for example with one layer of
perfect graphene plus a layer with a nh, such as that described above
Sect.~\ref{nanohole:sec}.
Figure~\ref{carb6_2lay:fig} shows this configuration with the insertion of
a C$_6$ polyyne.
The upper graphene layer and the polyyne are fully relaxed, while the lower
layer is kept frozen in ideal graphitic positions.
The equilibrium distance between the two layers, $332$~pm, is very close to
the observed interlayer distance of graphite, $335$~pm.

The resulting system composed of over $200$ atoms is computationally quite
expensive: indeed, with a modern parallel computer using $32$ Xeon-class
processors, it took more than two weeks to obtain a fully relaxed
configuration.
On the other hand, we verified that the properties of the upper layer do
not change significantly without the lower (perfect) sheet, whose effective
corrugation is quite small.
Indeed, the forces between the two layers are weak long-range forces whose
action is mild and almost translationally invariant, compared to the
intra-layer forces acting in the nh layer.
The DFT-LSDA evaluation of such weak dispersive forces is unreliable
anyway.
These observations suggest that it makes sense to consider the single layer
containing the nh and the C$_n$ spCC inserted into it, and leave the fixed
substrate layer out.
The relaxation of the positions, performed in the same conditions as above,
took about one week only in this 1-layer configuration involving $90$ C
atoms rather than 200 ones.

\begin{figure}
\begin{center}
\subfigure[nh-C$_5$]{\includegraphics[width=0.33\textwidth,angle=0,clip=]{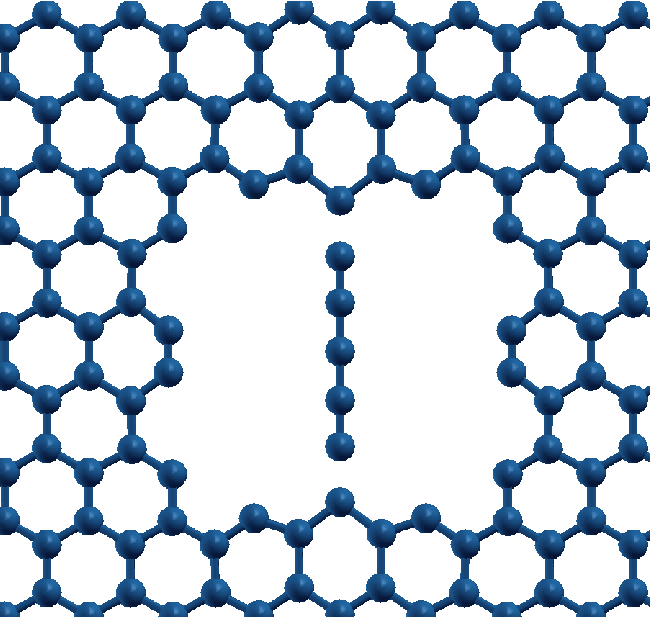}\label{carb5:fig}
  }
  \hspace{0.03\textwidth}
  \subfigure[nh-C$_5$ {\it $1$b}]{\includegraphics[width=0.33\textwidth,angle=0,clip=]{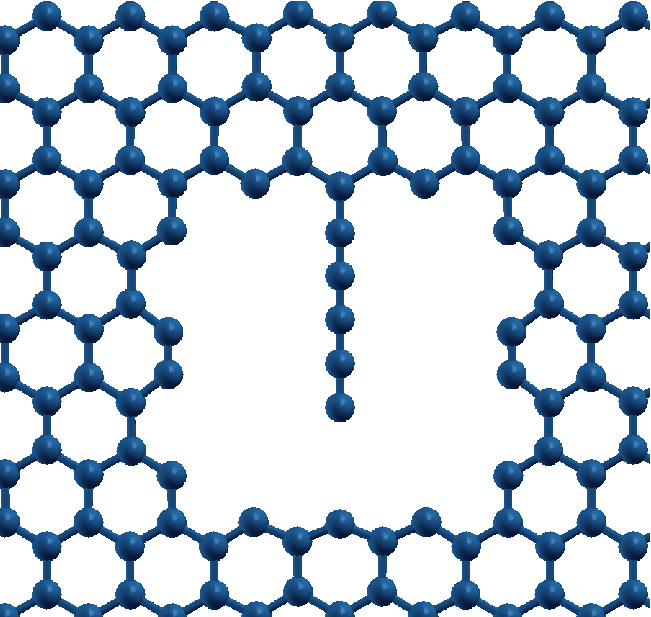}\label{carb5asi:fig}  }
  \hspace{0.03\textwidth}
  \subfigure[nh-C$_6$ {\it zig}]{\includegraphics[width=0.33\textwidth,angle=0,clip=]{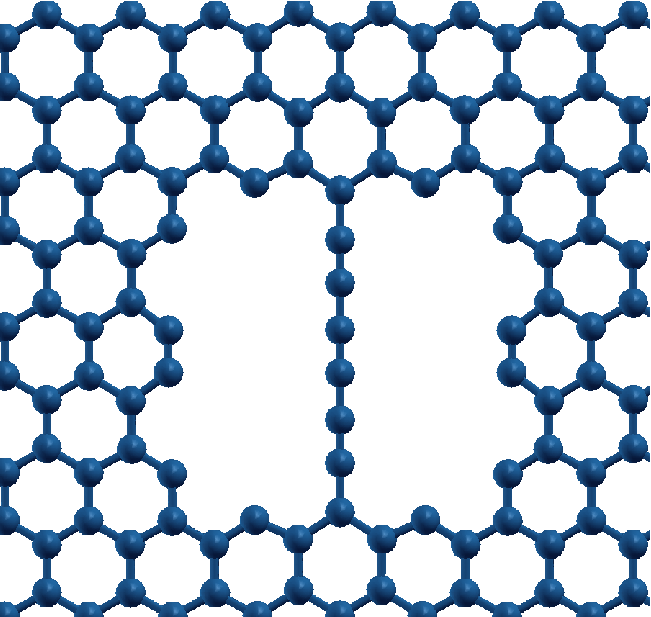}\label{carb6zig:fig}  }
  \hspace{0.03\textwidth}
  \subfigure[nh-C$_6$ {\it arm}]{\includegraphics[width=0.33\textwidth,angle=0,clip=]{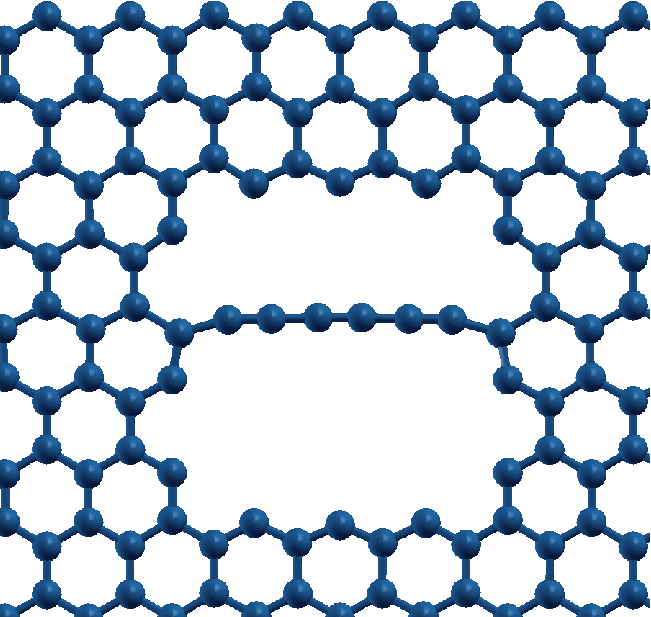}\label{carb6arm:fig}  }
  \hspace{0.03\textwidth}
  \subfigure[nh-C$_7$ {\it straight}, top view]{\includegraphics[width=0.33\textwidth,angle=0,clip=]{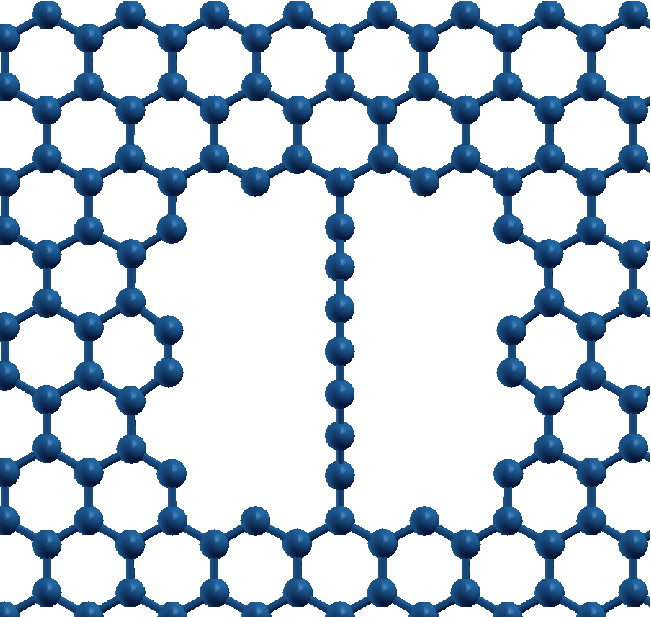}\label{carb7:fig}  }
  \hspace{0.03\textwidth}
  \subfigure[nh-C$_8$]{\includegraphics[width=0.33\textwidth,angle=0,clip=]{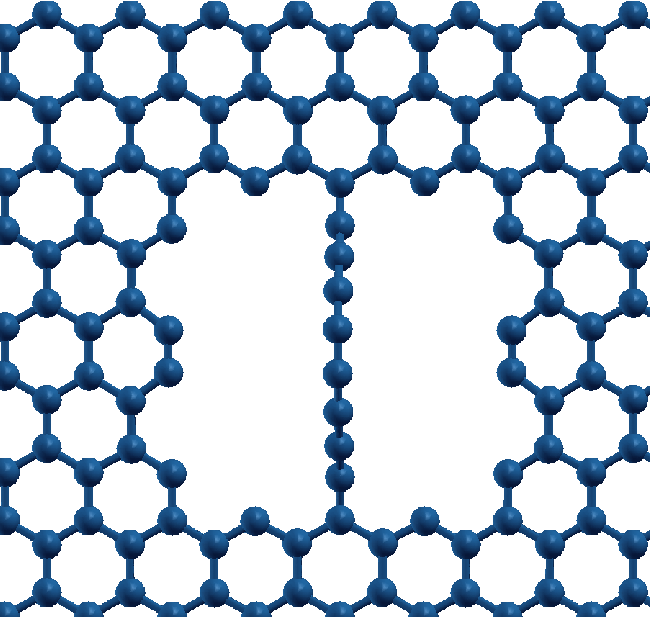}\label{carb8:fig}  }
  \hspace{0.03\textwidth}
  \subfigure[nh-C$_7$ {\it curved}, side view]{\includegraphics[width=0.33\textwidth,angle=0,clip=]{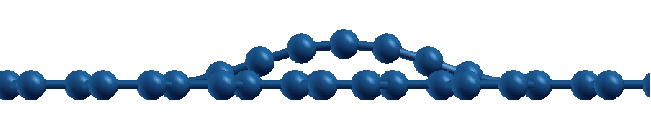}\label{carb7side:fig}  }
  \hspace{0.03\textwidth}
  \subfigure[nh-C$_7$ {\it s-curved}, side view]{\includegraphics[width=0.33\textwidth,angle=0,clip=]{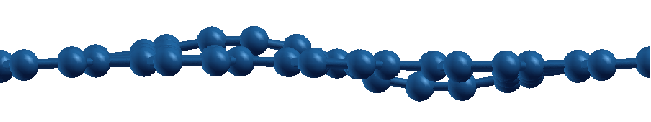}\label{carb7ondaside:fig} }
  \hspace{0.03\textwidth}
  \subfigure[nh-C$_7$ {\it straight}, side view]{\includegraphics[width=0.33\textwidth,angle=0,clip=]{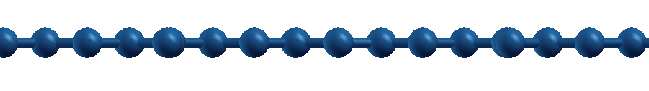}\label{carb7lineside:fig} }
  \hspace{0.03\textwidth}
  \subfigure[nh-C$_8$ side view]{\includegraphics[width=0.33\textwidth,angle=0,clip=]{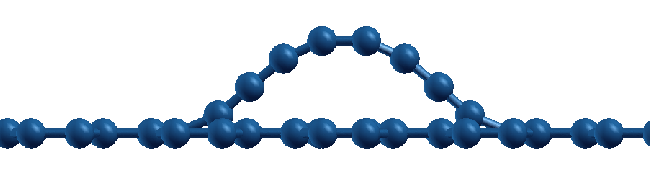}\label{carb8side:fig} }
  \hspace{0.03\textwidth}
  \subfigure{\includegraphics[width=1mm,angle=0,clip=]{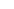}\scalebar{0.396\textwidth}{5}{3}{0}{1.5}{nm}}  
\end{center}
\caption{\label{carb_pos:fig} (Color online)
  The relaxed configurations of the spCC-nanohole structures considered in
  the present paper and described in Table~\ref{graph_carb:tab}.
}
\end{figure}

\begin{table}
\begin{center}
\begin {tabular}{p{0.19\textwidth}|p{0.52\textwidth}|c|c|c|c|c}
\hline
\hline
Label & Description & BLA & $E_{bond}$ &$\sigma_p$ &$\sigma_v$ &$\sigma_h$\\
\ [Fig.]&		    &[pm] & [eV]       & & &\\\hline\hline
nh-C$_5$\newline[Fig.~\ref{carb5:fig}] &
A C$_5$ chain stretched across the nh bonding weakly to both zig-zag edges
(bond lengths: $170$~pm).
& $7$ & $\ 4.2$ &\checkmark &\checkmark &\checkmark\\
nh-C$_5$ {\it $1$b}\newline
[Fig.~\ref{carb5asi:fig}]&
The C$_5$ chain bonded to one zig-zag edge of the nh; all bond lengths of
the spCC are similar to the lengths of cumulenic double bonds
($\simeq127$~pm).
& $1$ & $\ 6.2$ &\checkmark &\checkmark & \\\hline
nh-C$_6$ {\it  zig}\newline[Fig.~\ref{carb6zig:fig}]&
A weakly stretched C$_6$ chain bonded to the zig-zag nh edge.
& $10$ & $12.9$ &\checkmark &\checkmark &\checkmark\\\hline
nh-C$_6$ {\it arm}\newline[Fig.~\ref{carb6arm:fig}]&
The C$_6$ chain joining opposite armchair nh edges.
& $12$ & $\ 8.2$ &\checkmark &\checkmark & \\\hline
nh-C$_7$ {\it curved}\newline[Fig.~\ref{carb7side:fig}] &
A C$_7$ connected to the zig-zag edges, and buckling out of the graphene
plane (maximum spCC height $\simeq 120$~pm).
&$3$ &$12.0$&  &\checkmark &\checkmark\\
nh-C$_7$ {\it s-curved}\newline[Fig.~\ref{carb7ondaside:fig}]&
A C$_7$ connected to the zig-zag edges, buckling in a s shape, with the
central atom in the same plane as graphene.
&$3$& $11.9$ &* &\checkmark &*\\
nh-C$_7$ {\it straight}\newline[Figs.~\ref{carb7:fig},~\ref{carb7lineside:fig}]&
A compressed straight C$_7$ joining the zig-zag edges.
&$2$ &$11.8$ &\checkmark &\checkmark &\checkmark\\\hline
nh-C$_8$\newline [Figs.~\ref{carb8:fig},~\ref{carb8side:fig}] &
A C$_8$ curved chain. The maximum height of the spCC equals $297$~pm.
& $6$ & $12.5$ &  &\checkmark &\checkmark\\\hline
wnh-$2$C$_6$ \newline[Fig.~\ref{2carb6:fig}]\newline &
Two C$_6$ spCCs inserted in a wider nh joining the zig-zag edges.
The lateral distance between the spCCs equals $491$~pm.
& $11$ & n.c. &\checkmark &\checkmark &\checkmark\\
\hline \hline
\end{tabular}
\end{center}
\caption{\label{graph_carb:tab} (Color online)
  Summary of the individual configurations considered for the spCCs
  bound to the nh.
  For the relaxed configurations we report the resulting BLA and total
  bonding energy $E_{bond}$, corresponding to the formation of the (usually
  two) bonds between the spCC and the nh.
  The relevant symmetry planes ($\sigma_p$ is the reflection across the
  graphene layer plane, $\sigma_v$ is the vertical plane through the
  spCC, $\sigma_h$ is the horizontal plane through the middle of the
  spCC) are marked for the configurations for which they apply.
\newline
  $^*$Individual $\sigma_p$ and $\sigma_h$ are not symmetries for nh-C$_7$
  {\it s-curved}, but their product $\sigma_p \sigma_h$ is.
}
\end{table}

In our calculations, we consider several spCCs, from C$_5$ to C$_8$,
placed in different positions inside the nh.
We identify such compounds as nh-C$_n$, with further specification when
different local minima are considered.
Their relaxed configurations are depicted in Fig.~\ref{carb_pos:fig}.
Table~\ref{graph_carb:tab} summarizes the structural properties of the
configurations considered, comparing in particular their bond-length
alternation (BLA) \cite{BLAdefinition:note} and the spCC-graphene bonding
energy, defined as the total energy of the empty nh plus that of the
isolated spCC minus the total energy of the bonded spCC-nh configuration
under consideration.

For the selected nh size, C$_n$ chains of different length can fit more
or less easily inside the nh.
Short chains such as C$_5$ or C$_6$ may fit at the price of a tensile
stress, while longer spCCs can be forced inside the nh with a compressive
strain, which could be eased by buckling
\cite{Hu11,Castelli11,Cahangirov10}.

Strain influences directly the spCC BLA: a tensile strain leads to
stretching more the weaker bonds, thus producing an enhanced BLA, typical
of polyynic spCC (in nh-C$_6$ {\it arm}, whose BLA reaches $12$~pm).
Likewise, a BLA\,$\simeq 10$~pm is obtained for nh-C$_6$ {\it zig} due to the
nh size being approximately $5\%$ larger than the equilibrium length of the
spCC.
In contrast, a compressive strain leads to a more cumulenic-type structure,
e.g. nh-C$_7$ {\it straight} has BLA\;$\simeq 2$~pm.
As was observed in a slightly different context \cite{Hu11}, very small BLA
variations are induced by lateral atomic displacements.

The linear size of the hole is approximately $95\%$ of the equilibrium
length of the C$_7$ chain: different stable shapes of the spCC in
nh-C$_7$ can be stabilized by the compressive strain \cite{Castelli11}.
We study three equilibrium geometries of the spCC: straight,
single-curvature buckling, and s-curved buckling.
The compressive strain depresses the BLA, so that for the three of them the
BLA ranges from $2$ to $3$~pm.

The nh-C$_5$ is so much stretched that if kept in a central symmetric
configuration bonding between the spCC and the two edges of the nh is
weak, each highly stretched terminal bond contributing only about
$\simeq-2$~eV to lowering the total energy.
In such a condition we observe an intermediate BLA\;$\simeq 7$~pm.
This configuration is locally stable, but if we displace the spCC
significantly ($\simeq 50$~pm) closer to one nh edge than to the other, and
then let it relax, we retrieve an energetically favored configuration
(nh-C$_5$ {\it $1$b}) with essentially a single strong bond (total energy
lowering: $\simeq -6$~eV) between the spCC and the nh.
Here the spCC internal bond lengths are practically equal to those of
isolated C$_5$.

Due to the small size of the nh, the C$_8$ chains can only fit in a curved
geometry: the maximum out-of-plane elevation of the spCC equals $297$~pm.
The resulting BLA\;$\simeq 6$~pm is intermediate between cumulenic and
polyynic.

\begin{figure}
\centerline{
\includegraphics[width=0.6\textwidth,angle=0,clip=]{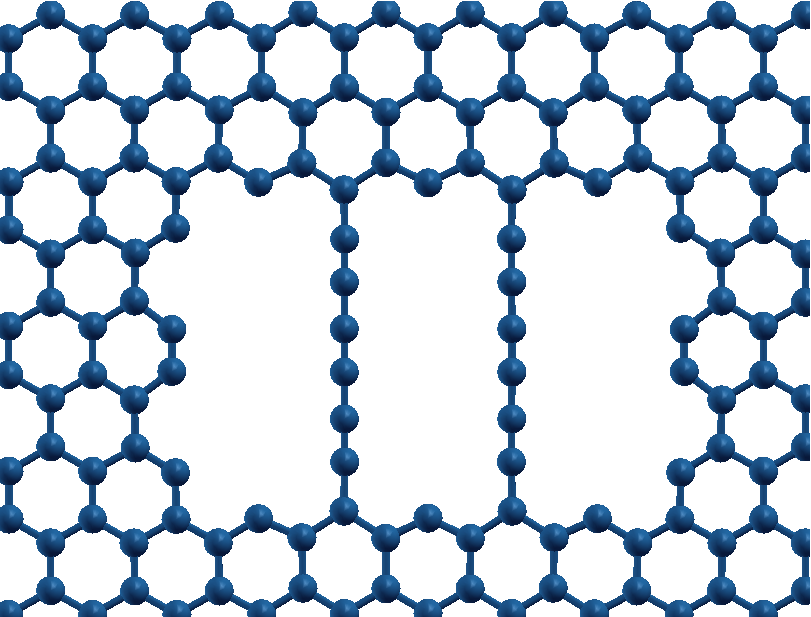}
}
\caption{\label{2carb6:fig} (Color online)
  Relaxed position of two C$_6$ spCCs in the wider nh: the hole width is
  1504\,pm, large enough to prevent the recombination of the two spCCs.
  This configuration displays no significant novelty relative to the single
  polyyne nh-C$_6$ {\it zig} configuration.
}
\end{figure}

We also consider a wider nh in which one can insert more than one spCC: in
wnh-$2$C$_6$, the nanohole contains two C$_6$ spCCs at a distance large
enough to keep them separated, see Fig.~\ref{2carb6:fig}.
All properties are essentially equivalent to those of the nh-C$_6$ {\it
  zig}, therefore we will not further investigate this configuration.

We evaluate the bonding energy of the configurations described here. Due to
its stretching, the nh-C$_5$ has a little value of $E_{bond} \simeq 4$~eV,
while for the nh-C$_5$ {\it $1$b} $E_{bond}=6.2$~eV which can be considered
a fair estimate of the spCC-graphene edge binding energy according to
DFT-LSDA, and matches previous evaluations \cite{Ravagnan09}.
For all other configurations $E_{bond} \simeq 12$~eV indicative of the
formation of two bonds, at the expense of approximately $1$~eV which
accounts for the elastic deformation energy of the spCC and the
connected graphene.

Until now the spCC was always connected to zig-zag edges.
When a C$_6$ chain binds to the armchair edges (nh-C$_6$ {\it arm}), the
bonding energy is smaller ($E_{bond} \simeq 9$~eV), due to the lower
reactivity of the armchair edge relative to the zig-zag one
\cite{Ravagnan09,Okada08}.
The BLA assumes a highly dimerized value $12$~pm associated to a tensile
strain, like for the nh-C$_6$ {\it zig} isomer.

In the following we shall investigate the electronic properties of selected
configurations.
In particular, we first focus on the interplay of the magnetic behavior of
the nh zig zag edges and of the spCC.
We will then move on to describe the DFT-LSDA band structure, the
vibrational properties, and the high-temperature stability of nh-C$_n$
configurations.

\section{Magnetism}\label{mag_graph:sec}

\begin{figure}
\centerline{
\includegraphics[width=0.65\textwidth,angle=0,clip=]{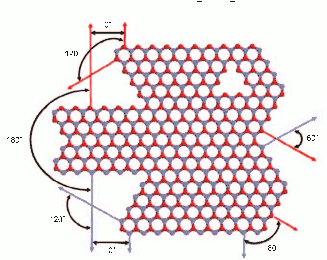}
}
\caption{\label{angle_graph:fig} (Color online)
  Scheme of the geometric relation between adjacent graphene edges.
  The angle between edges is defined as the angle between the vectors
  normal to the edge.
  The edge atoms belong to the same sublattice (either dark/red or
  clear/gray) when the zig-zag edges are at a relative angle of $0^{\circ}$
  or $120^{\circ}$; they instead belong to different sublattice when the
  relative angle is $60^{\circ}$ or $180^{\circ}$.
  (Adapted from Ref.~\cite{Yu08}.)
}
\end{figure}

Zig-zag edges are generally known to be ferrimagnetic
\cite{Yazyev08,Fujita96} due to non-bonding localized edge states.
A detailed investigation of the magnetic properties of graphene edge in the
context of a nanohole was carried out by Yu {\it et al.}~\cite{Yu08}.
That work focused on zig-zag edges (armchair ones are known to be
nonmagnetic \cite{Kusakabe03}), which made it convenient to study diamond-
or hexagon-shaped holes with zig-zag edges only.
The main conclusion of Ref.~\cite{Yu08} regarding consecutive zig-zag edges
is that the relative alignment of magnetic moments tends to be
ferromagnetic when the edge atoms belong to the same graphene sublattice.
This conclusion can be rephrased in terms of the angle between the two
consecutive edges, which is defined as the angle between the in-plane
outward vectors {\em normal to the edges}, as illustrated schematically
Fig.~\ref{angle_graph:fig}.
Ferromagnetic correlations occur when subsequent zig-zag edges are
unrotated ($0^{\circ}$) or rotated by $120^{\circ}$, as would happen in a
triangular hole.
In the opposite case, the magnetization is antiferromagnetic, as occurs
for zig-zag edges rotated by $60^{\circ}$ or $180^{\circ}$ (relevant,
e.g.\ for a diamond or and hexagonal hole).

\begin{figure}
\begin{center}
  \subfigure[nh-C$_7$]{\includegraphics[width=0.4\textwidth,angle=0,clip=]{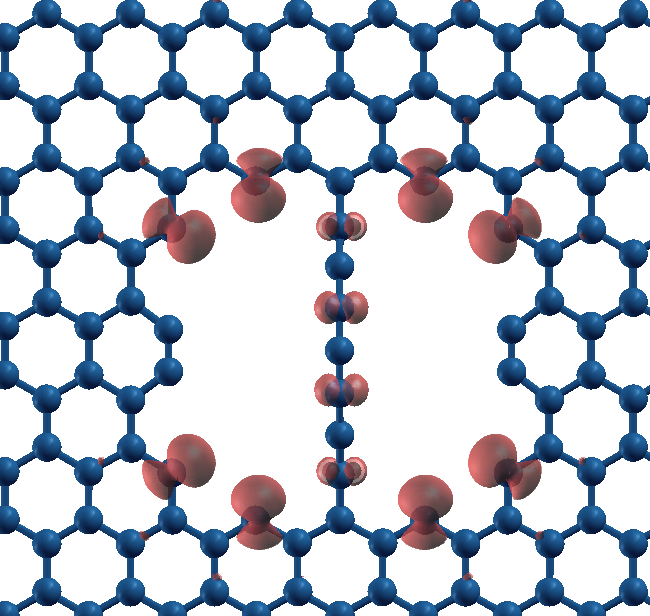}\label{graphite_carb7_ferro:fig}
  }
  \hspace{0.03\textwidth}
  \subfigure[nh-C$_8$]{\includegraphics[width=0.4\textwidth,angle=0,clip=]{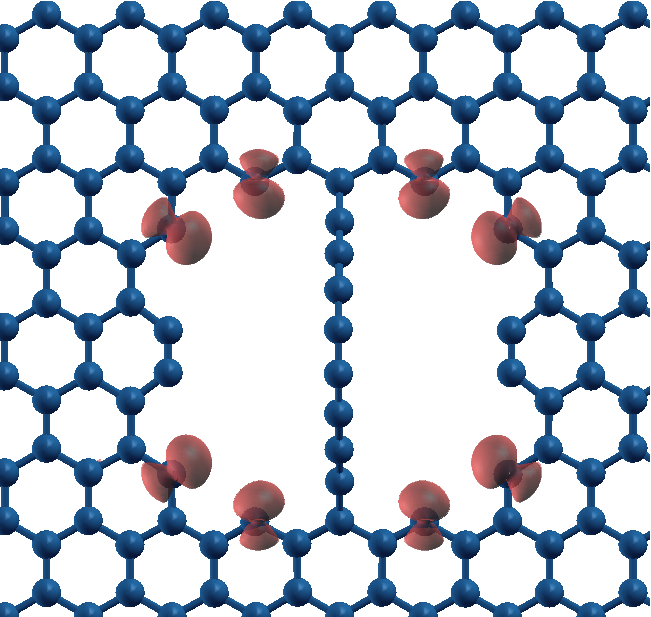}\label{carb8_fermag:fig}}
\end{center}
\caption{\label{carb_mag:fig} (Color online)
  The $0.01\,\mu_{\rm B}/a_0^3$ magnetic isosurface of the ferromagnetic
  state of nh-C$_7$ and of nh-C$_8$.
}
\end{figure}

Our rectangular nh involves two armchair edges, which are long enough to
isolate rather effectively the magnetic moments localized at the two zig-zag
edges.
In \ref{mag_nh_edge:sec}, we study the magnetic properties of the edge of
this nh.
Following spCC insertion, all structures described in
Sect.~\ref{carb_bind:sec} preserve a nonzero absolute magnetization,
associated to unpaired-spin electrons localized at the zig-zag nh edges.
A significant magnetization is shared by the C$_n$ spCCs with odd $n$,
while the even-$n$ spCCs are non-magnetic, as illustrated by
Fig.~\ref{carb_mag:fig} for the nh-C$_7$ and nh-C$_8$ structures.

\begin{figure}
\centerline{
\includegraphics[width=0.50\textwidth,angle=0,clip=]{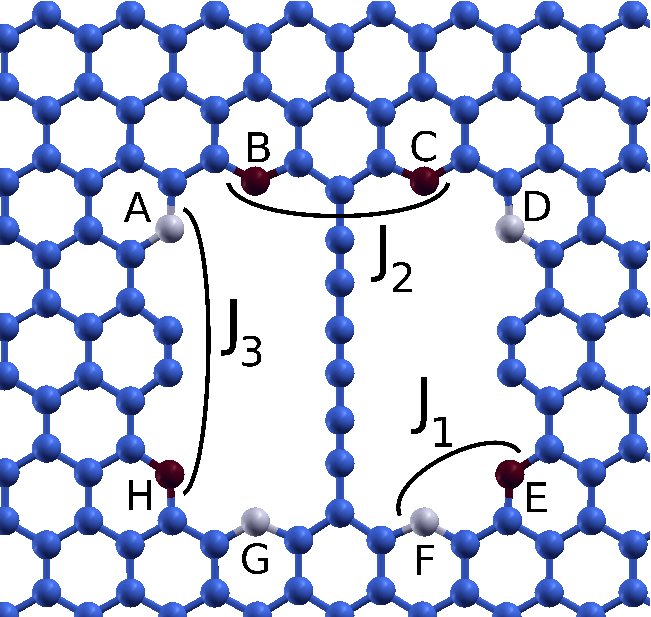}
}
\caption{\label{j_carb6:fig} (Color online)
  To investigate nonferromagnetic structures we consider two fictitious
  atomic carbon species fixing the spin polarization at the initial stage
  of the self-consistent calculation.
  C$_\uparrow$ (dark/red) and C$_\downarrow$ (clear/gray) carry positive
  and negative initial magnetization, respectively.
  Arcs mark all symmetry-independent nearest-neighbor Ising-type magnetic
  couplings $J_{ij}$, see Eq.~(\ref{ising_model}).
}
\end{figure}

The ferromagnetic structures of Fig.~\ref{carb_mag:fig} are induced by the
choice of a uniform starting magnetization used to initialize the
electronic self-consistent calculation.
To investigate other possible magnetic arrangements, we need to start off
the self-consistent calculation with different magnetic arrangements of the
individual atoms.
To do this, we define two fictitiously different atomic species, both with
the same chemical nature of C, but with initial magnetizations of opposite
sign ($\pm 1$~Bohr magneton).
We place these initially magnetically polarized atoms along the zig-zag
edges in order to trigger the desired magnetic structure.
Figure~\ref{j_carb6:fig} illustrates one of many possible arrangements of
the C$_\uparrow$ (dark/red) and C$_\downarrow$ (clear/gray) atoms to
initiate the self-consistent electronic-structure calculation.

\subsection{C$_6$-nh}\label{mag_c6nh:sec}

\begin{figure}
\begin{center} 
  \subfigure[Ground state: $E_{\rm tot}=$\newline$E_{\rm gs}=-13948.038$~eV\newline$M_{\rm tot}=0.00$\newline$M_{\rm abs}=9.31$]{\includegraphics[width=0.29\textwidth,angle=0,clip=]{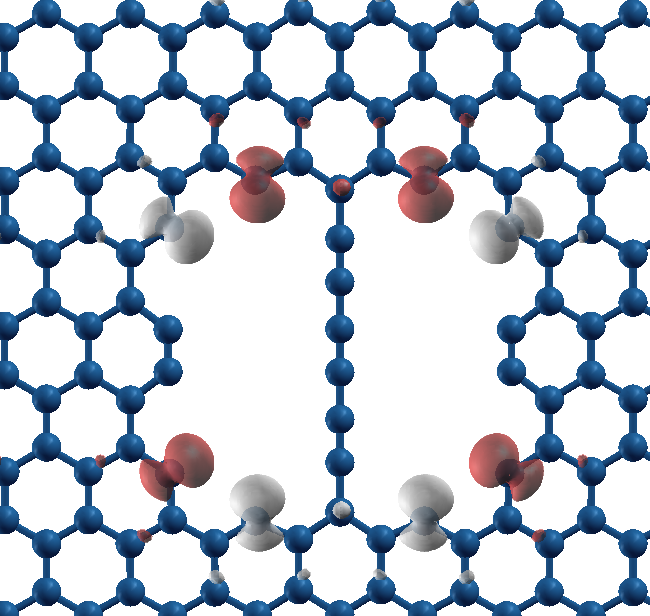}\label{carb6_mag1:fig}}\\
  \hspace{0.03\textwidth}
  \subfigure[$E_{\rm tot}=E_{\rm gs}+17$~meV\newline$M_{\rm tot}=0.00$\newline$M_{\rm abs}=8.21$]{\includegraphics[width=0.29\textwidth,angle=0,clip=]{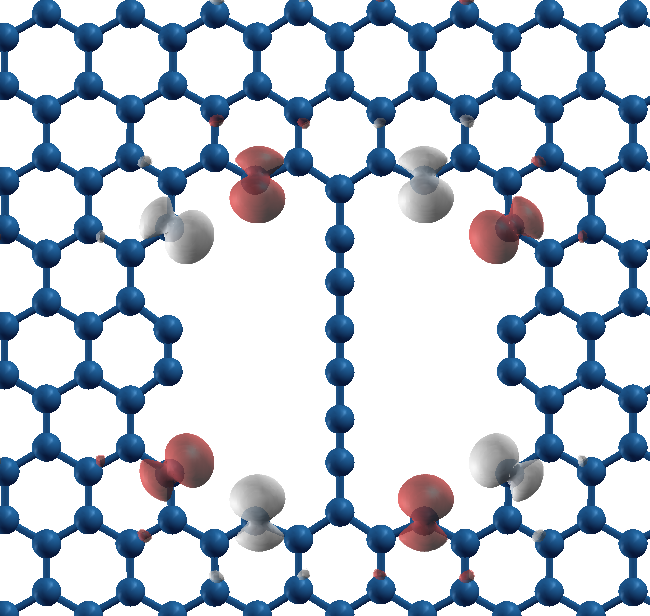}\label{carb6_mag5:fig}}
  \hspace{0.03\textwidth}
  \subfigure[$E_{\rm tot}=E_{\rm gs}+26$~meV\newline$M_{\rm tot}=0.00$\newline$M_{\rm abs}=8.03$]{\includegraphics[width=0.29\textwidth,angle=0,clip=]{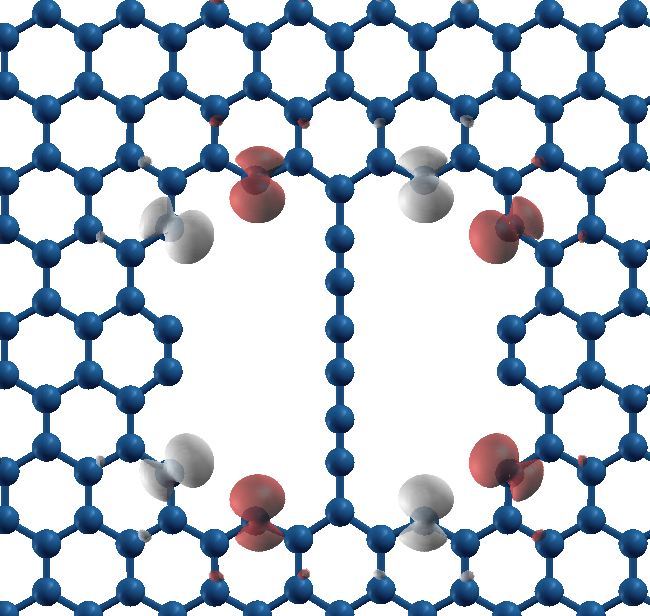}\label{carb6_mag6:fig}}
  \hspace{0.03\textwidth}
  \subfigure[$E_{\rm tot}=E_{\rm gs}+43$~meV\newline$M_{\rm tot}=0.00$\newline$M_{\rm abs}=8.38$]{\includegraphics[width=0.29\textwidth,angle=0,clip=]{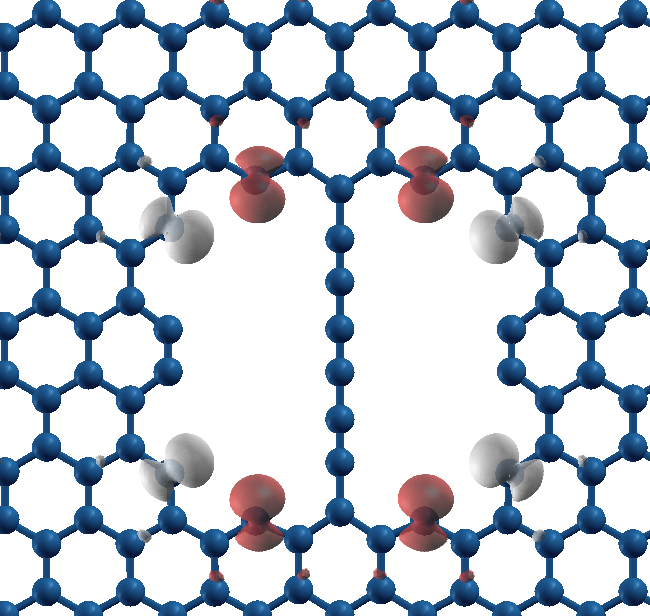}\label{carb6_mag4:fig}}
  \hspace{0.03\textwidth}
  \subfigure[$E_{\rm tot}=E_{\rm gs}+542$~meV\newline$M_{\rm tot}=0.00$\newline$M_{\rm abs}=8.67$]{\includegraphics[width=0.29\textwidth,angle=0,clip=]{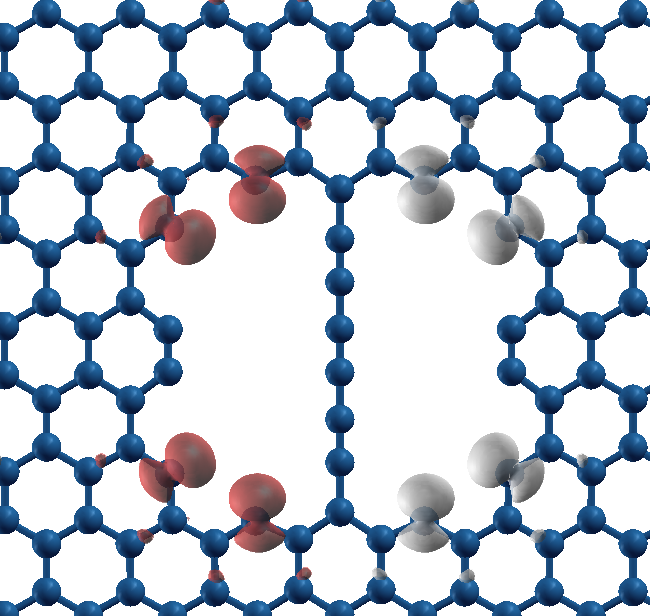}\label{carb6_mag2:fig}}
  \hspace{0.03\textwidth}
  \subfigure[$E_{\rm tot}=E_{\rm gs}+543$~meV\newline$M_{\rm tot}=0.00$\newline$M_{\rm abs}=9.19$]{\includegraphics[width=0.29\textwidth,angle=0,clip=]{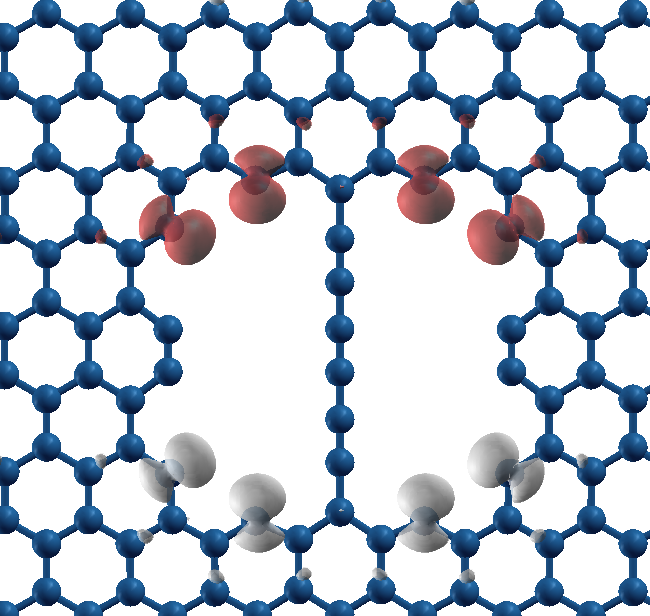}\label{carb6_mag3:fig}}
  \hspace{0.03\textwidth}
  \subfigure[$E_{\rm tot}=E_{\rm gs}+547$~meV\newline$M_{\rm tot}=8.00$\newline$M_{\rm abs}=8.66$]{\includegraphics[width=0.29\textwidth,angle=0,clip=]{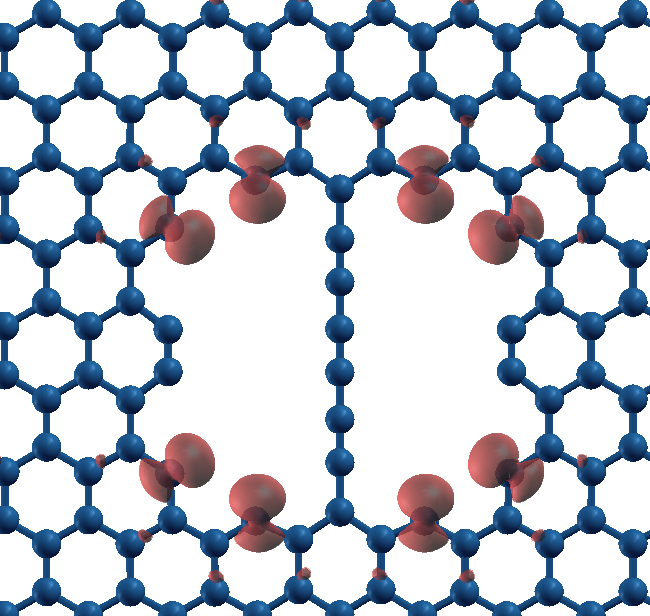}\label{carb6_mag7:fig}}
\end{center}
\caption{\label{carb6mag:fig} (Color online)
  Magnetization-density isosurfaces at $+0.01\,\mu_{\rm B}/a_0^3$ (dark/red)
  and $-0.01\,\mu_{\rm B}/a_0^3$ (clear/gray), for several different
  nh-C$_6$ magnetic structures. Frame (a): the DFT-LSDA ground state.
  Other frames: magnetically excited states.  For each frame we report the
  relevant excitation energy, total magnetization and integrated absolute
  magnetization in Bohr magnetons $\mu_{\rm B}$.
}
\end{figure}

We perform several self-consistent calculations for the nh-C$_6$ structure,
considering different starting magnetizations, as shown in
Fig.~\ref{carb6mag:fig}, and determine the ground magnetic configuration.
In agreement with Ref.~\cite{Yu08}, the ground-state configuration,
Fig.~\ref{carb6_mag1:fig}, has atoms of the the same magnetization in the
same graphene sublattice (e.g.\ atoms labeled B, C, H, E in
Fig.~\ref{j_carb6:fig}), and magnetization changes sign in passing from one
sublattice to the other.
The edge atoms bonded to even-$n$ C$_n$ spCCs show little magnetism,
mainly induced by the ferromagnetic interaction with neighboring atoms
along the same zig-zag edge.

This ground-state magnetic configuration is relaxed completely, and the
resulting total energy $E_{\rm gs}$ is taken as reference.
Keeping fixed this fully relaxed ground atomic configuration, we repeat
single self-consistent DFT-LSDA evaluations of the total energy, $E_{\rm
  tot}$, integrated magnetization $M_{\rm tot}= \int M_z(\vec r)\,d^3\vec
r$, and integrated absolute value of magnetization $M_{\rm abs}= \int
|M_z(\vec r)|\,d^3\vec r$, which we report next to each structure in
Fig.~\ref{carb6mag:fig}.
The resulting individual magnetic configurations, violating the
opposite-sublattice rule, are low-lying excitations, which we obtain in the
DFT-LSDA simulations by changing appropriately the initial magnetizations
of selected atoms.
Relaxation of one of these configurations shows very small displacements,
not larger than 6~pm.

The excitation energies of such states can be described approximately
within a Ising-model scheme.
The $z$ component $S_i$ of the spin degree of freedom accounting for the
magnetization at site $i$ interacts with neighboring spins $S_j$, with an
energy
\begin{equation}\label{ising_model}
H_{\rm spin}=-\sum_{<i , j>} J_{ij} S_i S_j
\,.
\end{equation}
According to the values of the absolute magnetization reported in
Fig.~\ref{carb6_mag1:fig}, it is appropriate to assume that each edge atom
carries one Bohr magneton, i.e.\ one unpaired spin $1/2$, thus $S_i=\pm
1/2$.
Accordingly, it makes sense to fit Ising-model energies only to
configurations with an absolute magnetization significantly close to
$8\,\mu_{\rm B}$.
To avoid parameter proliferation, we neglect interactions between
non-nearest-neighbor magnetic atoms.

Figure~\ref{j_carb6:fig} identifies the $3$ independent Ising interaction
parameters $J_k$ allowed by symmetry: $J_1$ for the interactions between
unpaired spins in different sublattices on edges rotated by $60^{\circ}$
($J_1=J_{AB}=J_{CD}=J_{EF}=J_{GH}$); $J_2$ for the interactions within the
same zig-zag edge, but ``isolated'' by the spCC ($J_2=J_{BC}=J_{FG}$);
and $J_3$ for the interactions across the armchair edge, representing
opposite sublattices, or edges rotated by $180^{\circ}$
($J_3=J_{AH}=J_{DE}$).
We write the energy of a configuration as the sum of the magnetic energy
$E_{\rm spin}$, approximated by the Ising expression (\ref{ising_model}),
plus $E_0$, including covalency and all other interactions establishing the
mean value of the total energy, averaged over all possible spin
orientations.
Specifically, for the nh-C$_6$ {\it zig} structure, we have:
\begin{eqnarray}\label{ising_carb6}
E_{\rm spin} = &-& J_1\left(S_A S_B+S_C S_D+S_E S_F+S_G S_H\right) \\\nonumber
&-&J_2\left(S_B S_C+S_F S_G\right) -J_3\left(S_A S_H + S_D S_E\right)
\,,
\end{eqnarray}
so that, given the ground configuration of Fig.~\ref{carb6_mag1:fig}, we
have $E_{\rm gs} = E_0 + E_{\rm spin} = E_0 + J_1 + (J_3-J_2)/2$.

\begin{table}
\begin{center}
\begin {tabular}{c|c|c}
\hline \hline Ising Parameter & Value [meV] & Standard deviation [meV]\\ \hline
$E_0-E_{\rm gs}$	& \ \ $281$ & $5$ \\
$J_1$ 			& $-259$ & $5$ \\
$J_2$ 			& \ \ $-4$ & $10$ \\
$J_3$ 			& \ $-17$ & $10$\\ \hline \hline
\end{tabular}
\end{center}
\caption{\label{j_carb6:tab} (Color online)
  The interaction parameters of the Ising Model, Eq.~\eqref{ising_model},
  fitted on the DFT-LSDA values of the total energy of the magnetic
  configurations of nh-C$_6$ {\it zig} listed in Fig.~\ref{carb6mag:fig}.
}
\end{table}

\begin{figure}
\centerline{
\includegraphics[width=0.75\textwidth,angle=0,clip=]{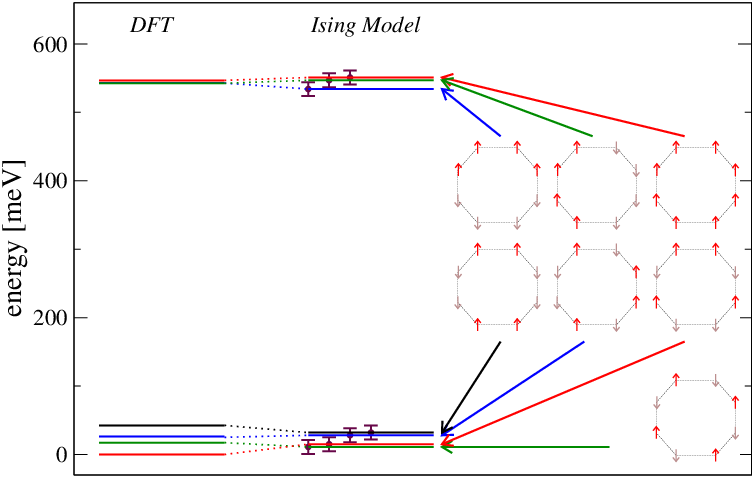}
}
\caption{\label{comp_carb6:fig} (Color online)
  Comparison of the DFT-LSDA energy levels of the magnetic structures of
  Fig.~\ref{carb6mag:fig} (left) with the spectrum (right) obtained using
  the Ising model, based on parameteres fitting the DFT values.
}
\end{figure}

We estimate the Ising-model parameters by means of a linear fit of the DFT
energies with expression (\ref{ising_carb6}).
Table~\ref{j_carb6:tab} reports the best-fit values of $E_0$ and $J_k$: all
the exchange energies $J_k$ turn out negative, reflecting antiferromagnetic
interactions.
The most significant value is $J_1$, reflecting the strong
antiferromagnetic coupling of adjacent unpaired spins on edge atoms
belonging to opposite sublattices.
$J_1$ is over one order of magnitude greater than the weakly
antiferromagnetic coupling $J_3$ across an armchair edge section.
Given the fit standard deviation, the small value of $J_2$ is compatible
with null coupling.
The obtained small negative value is the result of a strong cancellation
between the energy-order reversed configurations of panels
\ref{carb6_mag6:fig}, \ref{carb6_mag4:fig}, and the regularly ordered
states of panels \ref{carb6_mag1:fig}, \ref{carb6_mag5:fig}, the latter
matching the ordering Ref.~\cite{Yu08} as expected.
Indeed the Ising-model ground state and first-excited level turn out almost
degenerate and actually in reversed order due to the small positive value
of $J_2$, as illustrated in Fig.~\ref{comp_carb6:fig}.

In Fig.~\ref{comp_carb6:fig}, we compare the energies of the different
configurations of Fig.~\ref{carb6mag:fig} obtained by DFT calculation
with those obtained using the fitted Ising model.
A remarkable feature of the DFT excitation spectrum is the tiny splitting
of the levels of panels \ref{carb6_mag2:fig}-\ref{carb6_mag7:fig}, which is
hardly compatible with a simple nearest-neighbor Ising model.
Eventually Fig.~\ref{comp_carb6:fig} shows that the simple Ising model
fails to describe the fine structure of the magnetic excitation of the edge
atoms in the considered geometry.
Only the significant $J_1$ energy, fixing the rough structure of the
spectrum, is determined with fair accuracy.
Of course one could easily modify the model to include
e.g.\ second-neighbor interactions, to fit the detailed level structure,
but that would take any predictive power out of the model.

%

\subsection{C$_7$-nh}\label{mag_c7nh:sec}

\begin{figure}
\centerline{
  \subfigure[{\it Ground
      State}\newline$E_{tot}=E_{\rm
      gs}=-14102.247$~eV\newline$M_{tot}=-0.80$\newline$M_{abs}=8.97$]{\includegraphics[width=0.38\textwidth,angle=0,clip=]{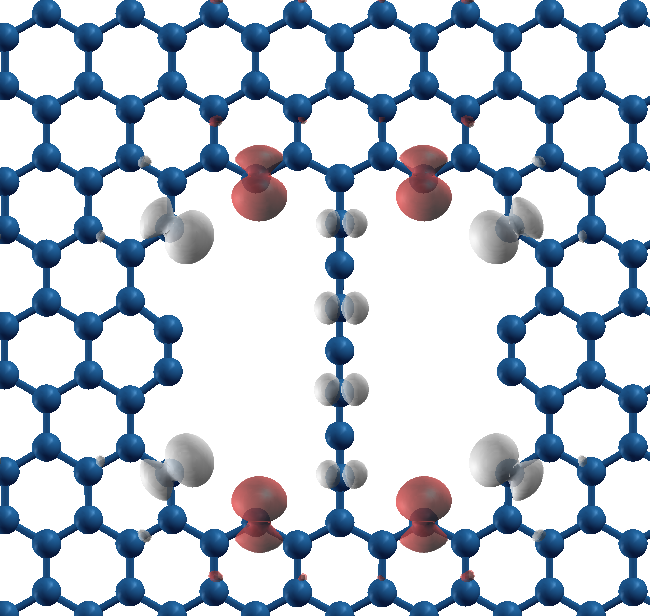}\label{carb7_mag1:fig}}
  \hspace{0.03\textwidth}
  \subfigure[$E_{tot}=E_{\rm
      gs}+60$~meV\newline$M_{tot}=-1.20$\newline$M_{abs}=9.63$]{\includegraphics[width=0.38\textwidth,angle=0,clip=]{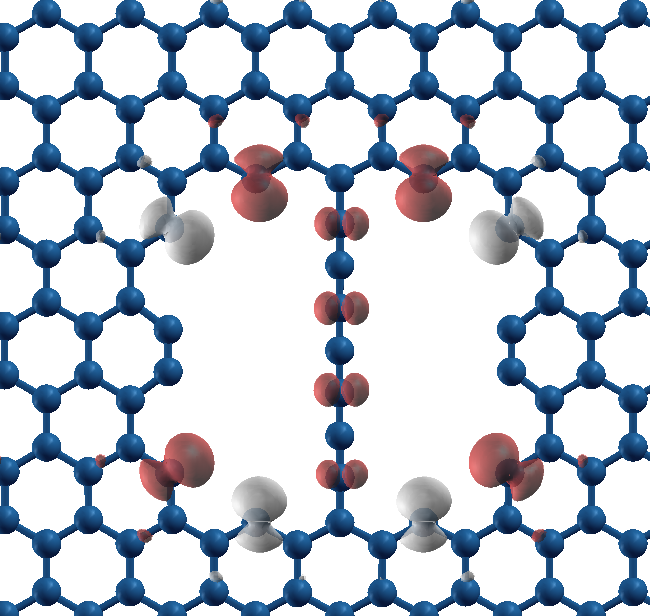}\label{carb7_mag2:fig}}
}
\caption{\label{carb7mag:fig} (Color online)
  Magnetization isosurfaces at $+0.01\,\mu_B/a_0^3$ (dark/red) and
  $-0.01\,\mu_B/a_0^3$ (clear gray) for two magnetic structures of nh-C$_7$
  {\it straight}.
}
\end{figure}

One may attempt a similar analysis for the odd-$n$ spCCs,
e.g.\ nh-C$_7$.
As Fig.~\ref{graphite_carb7_ferro:fig} shows, odd-$n$ spCCs are magnetic, thus
quite different from the even-$n$ ones.
This leads to two consequences for odd spCCs attached to nh: first the spin
values at different sites are different, and second the number of spin
interactions to be considered is greater.
This would leave little significance to a Ising model description.

It is possible to at least identify the ground magnetic configuration, like
we did for even-$n$ chains.
Figure~\ref{carb7mag:fig} shows two different magnetic configuration of the
nh-C$_7$ {\it straight} structure.
The ground-state configuration is the one of Fig.~\ref{carb7_mag1:fig},
which follows the edge rules of Ref.~\cite{Yu08}.
The coupling between the spCC and nh edges is antiferromagnetic: this can
be seen as a special case of the edge rules if the spCC atoms are seen as
graphene atoms belonging to the edge but in the other sublattice relative
to the outer zig-zag edge magnetic atoms.
This coupling is so strong that it prevails over the weak antiferro
long-range $J_3$-type coupling.
One can estimate this magnetic coupling energy between the end spCC atom
and one of the nearest zig-zag edge atoms to approximately $\approx
100$~meV.
The intra-spCC interaction is distinctly antiferromagnetic.

\section{Electronic Properties}\label{elect_graph:sec}

\begin{figure}
\centerline{
  \subfigure[{\bf k}-point path]{\includegraphics[width=0.25\textwidth,angle=0,clip=]{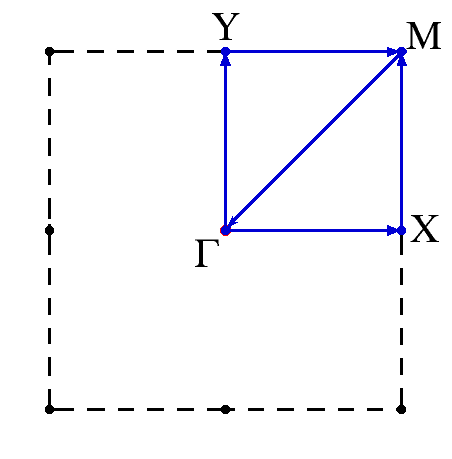}\label{kpoint:fig}}
  \hfill
  \subfigure[nh bands]{\includegraphics[width=0.65\textwidth,angle=0,clip=]{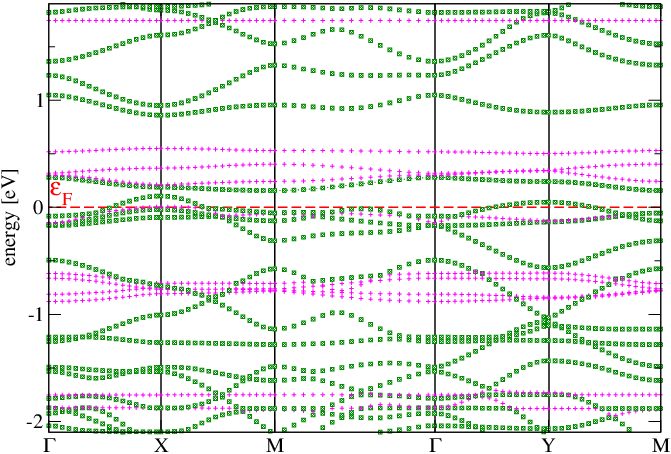}\label{bands_hole:fig}}
}
\caption{\label{band_hole:fig} (Color online)
  (a): The Brillouin-zone (dashed) with the $\Gamma-X-M-\Gamma-Y-M$ {\bf
    k}-point path adopted for all band-structure calculations of the
  present paper.
  (b) Spin-majority Kohn-Sham band structure of the relaxed nh superlattice
  of Fig.~\ref{hole_pos:fig}, in the ferromagnetic configuration of
  Fig.~\ref{hole_mag6:fig}: magenta crosses stand for HE bands localized on
  the hole-edge; green squares represent delocalized BU states.
  The plot focuses a $4$~eV-wide energy region around the Fermi level (red
  dashed) for better readability.
}
\end{figure}

The magnetic properties described in the previous section are determined by
the electronic structure.
Before analyzing the nanohole-spCC system, it is useful to examine the
simpler bands of a empty nh.
We will track the the bands along the Brillouin-zone path shown in
Fig.~\ref{kpoint:fig}.
We sample the ${\bf k}$-space path with points separated by
$1.75\times 10^{-13}$~m$^{-1}$.

\begin{figure}
\centerline{
  \subfigure[A HE state]{\includegraphics[width=0.4\textwidth,angle=0,clip=]{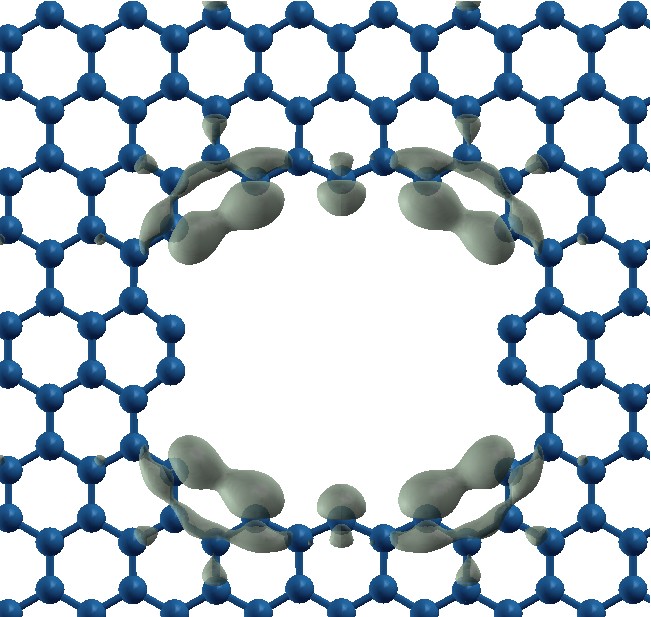}\label{hole_wave:fig}  }
  \hspace{0.03\textwidth}
  \subfigure[A BU state]{\includegraphics[width=0.4\textwidth,angle=0,clip=]{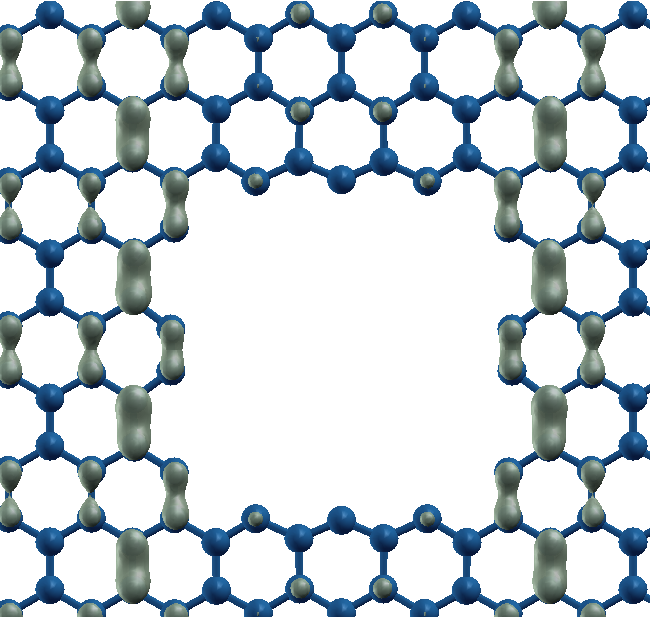}\label{mix_wave:fig}  }
  }
\caption{\label{wave_hole:fig} (Color online)
  Examples of ${\bf k}=0$ electronic states of the nh.
  (a): a localized HE state; (b): a BU state.
  In Fig.~\ref{bands_hole:fig}, these states are located at the
  $\Gamma$-point crossing at $-0.81$~eV and $1.36$~eV, respectively.
}
\end{figure}

Figure~\ref{bands_hole:fig} displays the band structure near the Fermi
energy for the empty nh superlattice of Fig.~\ref{hole_pos:fig}.
One can identify two kinds of bands, with different spatial localization
properties of their wave functions:
(i) States localized at the hole edge (HE), such as the one depicted in
Figs.~\ref{hole_wave:fig}.  We use magenta crosses to track these HE states
in the bands-structure plots, such as Fig.~\ref{bands_hole:fig}.
(ii) States like the one in Fig.~\ref{mix_wave:fig} localized primarily on
the bulk graphene atoms, with a weak component on the edge atoms.  We label
these states as BU, and identify them with green squares in band-structure
plots.

HE bands are generally flat, with little dispersion.
The small but nonzero bandwidth of HE states is due to the residual
interaction between the nh and its periodic images.
A HE band touches the Fermi energy near $X$, and is therefore only partly
filled, thus becoming the responsible of the edge magnetism discussed in
Sect.~\ref{mag_graph:sec}.

Liu {\it et al.}\ in their investigation of the band structures of a
different graphene nanohole \cite{Liu09}, discovered the opening of band
gaps for nanoholes with either armchair or zig-zag edges.
In contrast, our graphene with a isolated nh exhibits no band gap at the
Fermi energy, and retains the (semi)metallic character of graphene.
Specifically, a BU metallic band crosses $\epsilon_F$ and shows a modest
but distinct dispersion, with sizeable empty hole pockets near the $X$ and
$Y$ points.

\begin{figure}
\begin{center}
  \subfigure[nh-C$_5$]
            {\includegraphics[width=0.65\textwidth,angle=0,clip=]{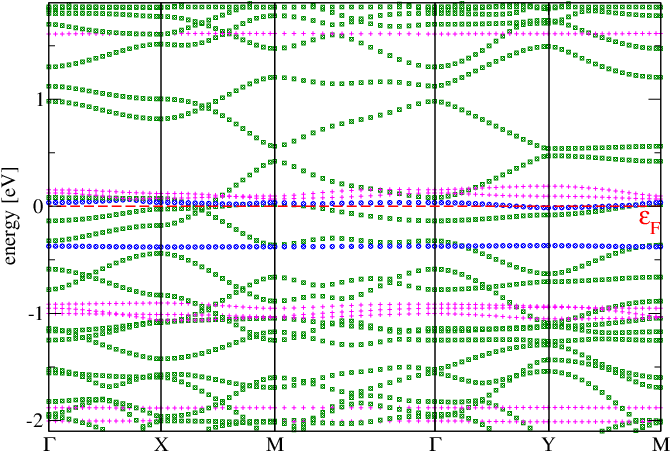}\label{bands_carb5:fig}}
  \hspace{0.01\textwidth}
  \subfigure[nh-C$_5$ {\it $1$b}]
            {\includegraphics[width=0.65\textwidth,angle=0,clip=]{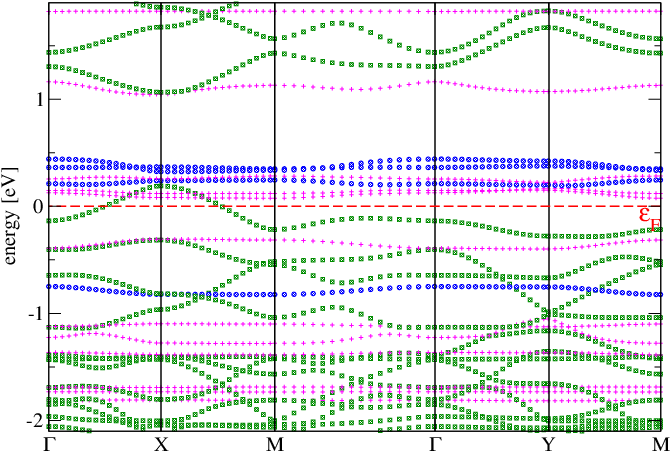}\label{bands_carb5asi:fig}}
\end{center}
\caption{\label{bands1_carb:fig} (Color online)
Spin-majority band structures of (a) nh-C$_5$, represented in
Fig.~\ref{carb5:fig}, and of (b) nh-C$_5$ {\it $1$b},
Fig.~\ref{carb5asi:fig}.
}
\end{figure}

\begin{figure}
\begin{center}
  \subfigure[nh-C$_6$ {\it zig} majority
    spin]{\includegraphics[width=0.65\textwidth,angle=0,clip=]{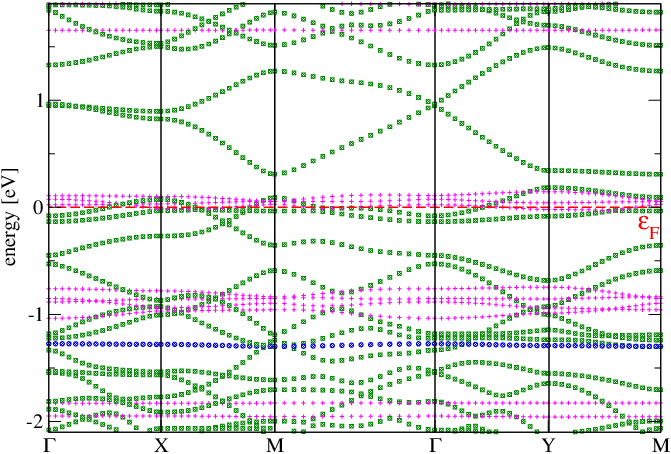}\label{bands_carb6up:fig}}
  \hspace{0.01\textwidth}
  \subfigure[nh-C$_6$ {\it zig} minority
    spin]{\includegraphics[width=0.65\textwidth,angle=0,clip=]{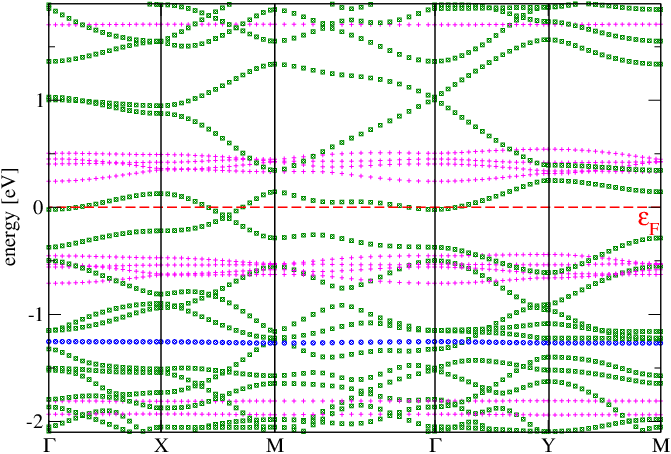}\label{bands_carb6down:fig}}
\end{center}
\caption{\label{bands2_carb:fig} (Color online)
  Band structure for the majority (a) and minority (b) spin components of
  the ferromagnetic state of nh-C$_6$ {\it zig}, depicted in
  Fig.~\ref{carb6zig:fig}.
}
\end{figure}

\begin{figure}
\begin{center}
  \subfigure[nh-C$_6$ {\it
      arm}]{\includegraphics[width=0.65\textwidth,angle=0,clip=]{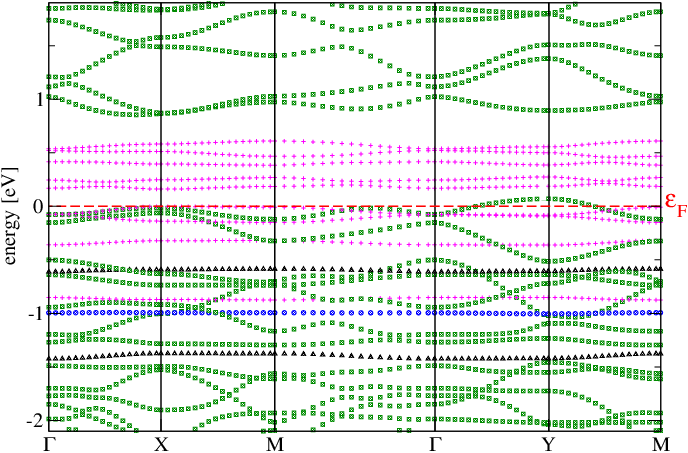}\label{bands_carb6arm:fig}}
  \hspace{0.01\textwidth}
  \subfigure[nh-C$_7$]{\includegraphics[width=0.65\textwidth,angle=0,clip=]{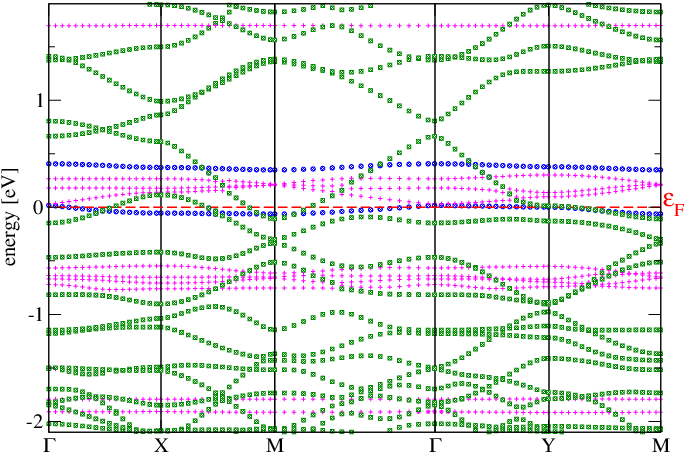}\label{bands_carb7:fig}}
\end{center}
\caption{\label{bands3_carb:fig} (Color online)
  Spin-majority band structures of (a) the nh-C$_6$ {\it arm}
  nanostructure, depicted in Fig.~\ref{carb6arm:fig}, and 
  (b) nh-C$_7$, Fig.~\ref{carb7:fig}.
}
\end{figure}

\begin{figure}
\centerline{
  \subfigure[nh-C$_8$]{\includegraphics[width=0.65\textwidth,angle=0,clip=]{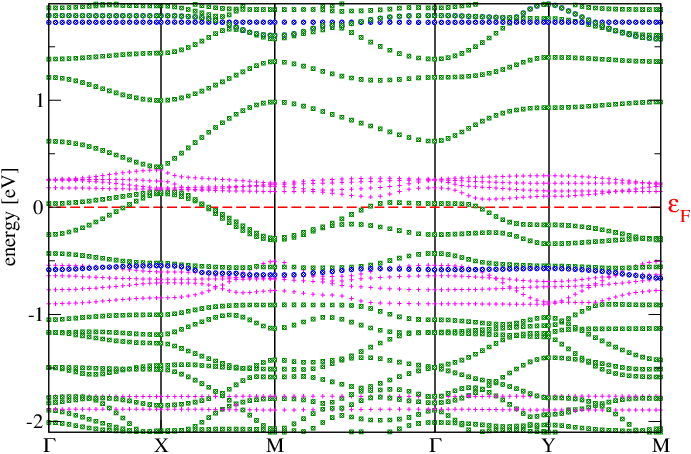}\label{bands_carb8:fig}
} }\centerline{
  \subfigure[nh-$2$C$_6$]{\includegraphics[width=0.65\textwidth,angle=0,clip=]{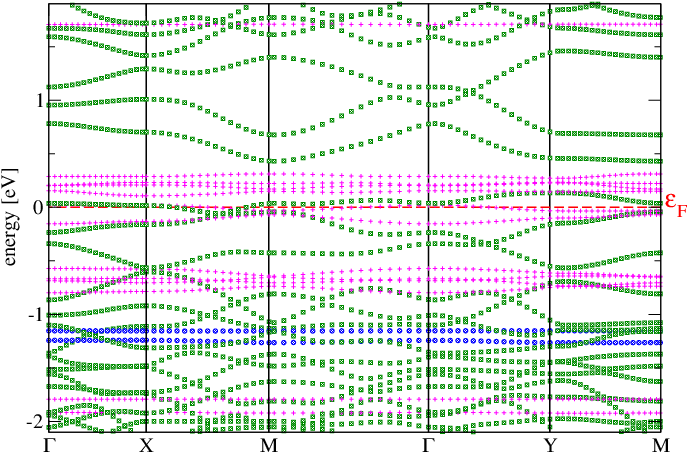}\label{bands_2carb6:fig}
} }
\caption{\label{bands4_carb:fig} (Color online)
  Spin-majority band structures of nh-C$_8$, see Fig.~\ref{carb8:fig}, and
  of nh-$2$C$_6$, see Fig.~\ref{2carb6:fig}.
}
\end{figure}

\begin{figure}
\begin{center}
  \subfigure[A CB state]
            {\includegraphics[width=0.3\textwidth,angle=0,clip=]{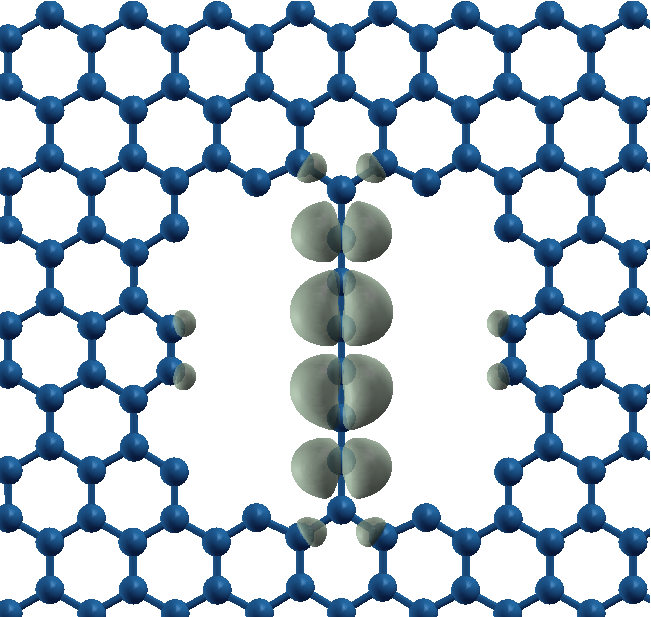}\label{carb_wave:fig}  }
  \hspace{0.01\textwidth}
  \subfigure[A CHE state]
            {\includegraphics[width=0.3\textwidth,angle=0,clip=]{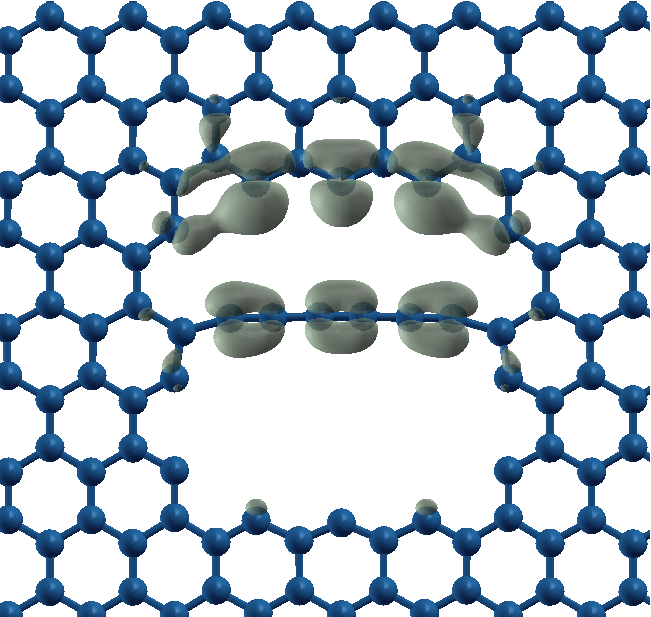}\label{mixcarb_wave:fig}  }
  \hspace{0.01\textwidth}
  \subfigure[A resonant bulk state]
            {\includegraphics[width=0.3\textwidth,angle=0,clip=]{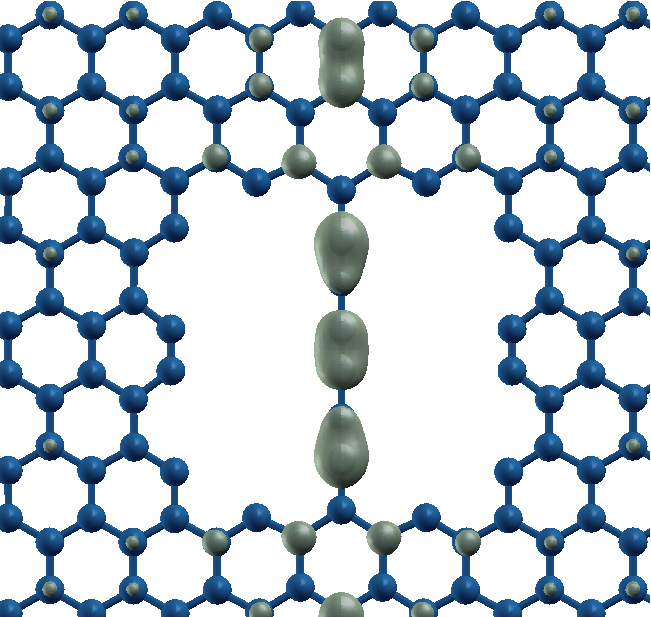}\label{BH_states:fig}  }
\end{center}
\caption{\label{wave_carb:fig} (Color online)
  Like in Fig.~\ref{wave_hole:fig}, but for nh-C$_6$ {\it zig} and nh-C$_6$
  {\it arm}.
  (a): a typical CB state localized mainly on the polyyne -- the $\Gamma$
  state at energy $-1.16$~eV in Fig.~\ref{bands_carb6up:fig};
  (b): a CHE state, localized jointly on the polyyne and the nh edge -- the
  $\Gamma$ state at energy $-0.50$~eV in Fig.~\ref{bands_carb6arm:fig};
  (c): a bulk state showing a significant extension on the polyyne -- the
  $\Gamma$ state at energy $-0.46$~eV in Fig.~\ref{bands_carb6up:fig}.
}
\end{figure}

Coming to the nh-C$_n$ systems, Figs.~\ref{bands1_carb:fig},
\ref{bands2_carb:fig}, \ref{bands3_carb:fig}, and \ref{bands4_carb:fig}
report details of the computed DFT-LSDA bands for the relaxed structures of
Figs.~\ref{carb_pos:fig} and \ref{2carb6:fig}.
Also in these band structures we identify HE (magenta crosses) and BU
(green squares) bands.
In addition, band states significantly localized on the spCC atoms (CB)
are identified by blue circles.
One such state is depicted in Fig.~\ref{carb_wave:fig}.
Occasional resonances of localized spCC and nh-edge states lead to
hybrid localized states involving both, e.g.\ the one of
Fig.~\ref{mixcarb_wave:fig}.
We identify such ``CHE'' states only in nh-C$_6$ {\it arm}, and label them
by black triangles in Fig.~\ref{bands_carb6arm:fig}.
Graphene bulk states often hybridize with the spCC molecular orbitals, thus
acquiring a significant spCC components, as illustrated for example in
Fig.~\ref{BH_states:fig}.
Our sample is too small to distinguish clearly between entirely localized
states at the spCC/nh edge (whose bands would be perfectly flat in a
realistically wide sample) and only partly localized hybrid states.

Essentially all structures display a metallic behavior, due to one or
several bulk bands crossing the Fermi energy.
The different bonded spCCs affect the graphene nh bands quite
considerably, by both shifting them and deforming them especially near the
Fermi energy.
In particular, the positions of several localized states at the nh edge
change depending on the spCC state, and moreover spCC-specific
localized states occur.
For even-$n$ spCCs, the CB states are energetically quite distant from the
Fermi level, while odd-$n$ spCCs exhibit a more metallic behavior, with
spCC states quite close to the Fermi energy, and significant
hybridization with the extended bulk states, consistently with results of Ref.~\cite{Ravagnan09,Zanolli11}.

The comparison of the spin majority and minority bands in
Fig.~\ref{bands2_carb:fig} shows that magnetism affects the bulk bands only
weakly.
Magnetism appears to be associated to an energy shift of a few localized HE
and (for odd spCCs) CB states near the Fermi level.
The resulting effective exchange energy is $\simeq 0.3$~eV.

We perform several calculations of the nh-C$_7$ structure for each of the
considered spCC shapes:
curved -- Fig.~\ref{carb7side:fig},
s-curved -- Fig.~\ref{carb7ondaside:fig}, and
straight -- Fig.~\ref{carb7lineside:fig}.
All these geometries show basically identical band structures, e.g.\ the
one reported in Fig.~\ref{bands_carb7:fig}.
Likewise, no special effect of the spCC curvature is apparent in the
bands of nh-C$_8$, Fig.~\ref{bands_carb8:fig}.
Finally, the congestion of the bands near the Fermi level in
Fig.~\ref{bands_2carb6:fig} is a consequence of the larger cell, and
greater number of atoms and of electrons of this specific configuration.
The general considerations (even-$n$ spCC bands away from the Fermi
energy, magnetism related to HE bands near the Fermi energy) apply also in
this more intricate configuration.

\section{Vibrational spectra}\label{vibr_graph:sec}

We perform phonon calculation for a few stable structures of
Sect.~\ref{carb_bind:sec}.
We evaluate the phonon frequencies and eigenvectors of using standard
density-functional perturbation theory, as implemented in the Quantum
Espresso code \cite{espresso2009,Baroni02}.
For comparison, the theoretical C-C stretching modes of polyynes
C$_{n}$H$_2$ ($n=8-12$) evaluated with the same method match the
experimental frequencies \cite{Tabata06} to within 40~cm$^{-1}$.
The size of the system is too large to evaluate the full dynamical matrix:
although in principle possible, it would require a huge investment of
computer time.

We focus specifically on spCC ``optical'' stretching modes, which are
prominent and characteristic in the experimental spectra of $sp - sp^2$
carbon in the spectral region near 2000~cm$^{-1}$, while all other
vibrations (the ``acoustic'' spCC stretching modes, all bending modes,
all graphene vibrations) overlap and lump together in a continuum extending
from 0 to 1600 cm$^{-1}$ \cite{Ravagnan02,Ravagnan09}.
We verified that the spCC stretching modes are influenced very little by
faraway ligand atoms \cite{Castelli10}.
Accordingly, we only compute and diagonalize the part of the dynamical
matrix at $\Gamma$ relative to displacements of the atoms of the spCC,
plus its first and second neighbors in the graphene sheet.
The error in the vibrational frequency induced by this approximation can be
estimated $\simeq 1$~cm$^{-1}$.

\begin{table}
\begin{center}
\begin {tabular}{c|c|c}
\hline \hline Structure name & Raman frequencies~[cm$^{-1}$]& IR
frequencies~[cm$^{-1}$]\\ \hline nh-C$_5$ & $1323$, $1368$ &
$1332$ \\ nh-C$_6$ {\it zig} & $1777$, $\mathbf{1878}$ &
$\mathbf{1939}$ \\ \hline \hline
\end{tabular}
\end{center}
\caption{\label{graph_freq:tab}
  Wavenumber of Raman and IR spCC stretching frequencies calculated for the
  nh-C$_n$ structures.
  The most intense Raman and IR frequencies are highlighted in {\bf bold}.
}
\end{table}

By analyzing the displacement pattern of the normal modes of the even-$n$
spCCs, it is straightforward to identify the ``$\alpha$'' modes
characterized by the strongest Raman and IR absorption
\cite{Ravagnan09,Cataldo10,Cinquanta11,Innocenti10}.
We take advantage of this pattern recognition to avoid a computationally
expensive explicit evaluation of the Raman and IR intensities.
Table~\ref{graph_freq:tab} reports the computed frequencies of the spCC
stretching modes, with the most intense Raman and IR mode highlighted.
Note that the wavenumbers of the frequencies are significantly lower than
the characteristic stretching-frequencies of free spCCs ($1950 -
2300~\rm{cm^{-1}}$).
The reason is the tensile strain to which the spCCs are subjected by
binding to the nh edges.
%
In nh-C$_5$, the length of the chain, including the bonds between the
spCC and the nanohole, is $15\%$ longer than the isolated spCC; this
elongation leads to frequencies much softer than typical polyyne ones.
The chain length in nh-C$_6$ {\it zig} is only $5\%$ longer than isolated
length, and the frequencies come much closer to the typical frequencies of
free spCCs.
%

The results of the present section do not imply that spCCs in a context
of nanostructured $sp$-$sp^2$ carbon material should vibrate at much
different frequencies from their molecular counterparts \cite{Cataldo10}.
Quite on the contrary, previous calculations and experiments confirm that
fully relaxed spCCs terminated by $sp^2$-type material exhibit very
similar frequencies to those of molecular spCCs
\cite{Ravagnan09,Cataldo10,Cinquanta11}.
The results of the present calculations suggest instead that unrelaxed
tensile strain in nanostructured $sp$ - $sp^2$ carbon material is likely to
induce significant frequency shifts of the spCC modes.
Depending on the method of production of spCC-containing material
(e.g.\ cluster beam formation/deposition \cite{Ravagnan02} vs.\ atomic
wires stretched out from pulled graphene sheets \cite{Jin09,Chuvilin09}),
whenever a sizeable tensile strains remain frozen in the sample, one is to
observe a corresponding distribution of the observed vibrational
frequencies, quite independent of the frequency shifts associated to
different lengths and terminations of the spCCs \cite{Cinquanta11}.

\section{High-temperature stability}\label{highT:sec}

All studied configurations are local minima of the adiabatic potential
energy, i.e. metastable allotropes of carbon.
Given sufficiently long time, the spCCs are expected to degrade, for
example by recombining with the nh edge and extend energetically favored
$sp^2$ graphene.
This possibility is however very remote at low temperature, because in this
interconversion process the energy barriers to be crossed are quite
substantial.

Since several detailed types of $sp \to sp^2$ processes are possible, a
full study of the spCC degradation is beyond the scope of the present
paper.
We content ourselves with a semi-quantitative estimate of the thermal
stability and the degradation mechanisms of spCCs bonded to carbon
$sp^2$ nanostructures by running comparably long high-temperature
molecular-dynamics (MD) simulations, and monitor the eventuality of spCC
decomposition as a function of the simulated temperature.

As, due to the size dependency of statistical fluctuations, the longer the
spCCs the higher is the chance of chain breaking.
We therefore prefer to simulate a larger version of the model of
Sect.~\ref{carb_bind:sec}, namely a C$_{10}$ chain bound to a nh large
enough for it to fit loosely.  Also the graphene plane is represented by
four, rather than three hexagonal rings separating the nh periodic
replicas.
The resulting nh-C$_{10}$ structure involves 184 atoms and 736 electrons.
Even using Car-Parrinello dynamics, it would be a formidable task to
simulate several sufficiently long runs to expect a significant chance to
observe spCC dissociation at a temperature comparable to experiment.
For the present task therefore we abandon the {\it ab-initio} DFT-LSDA
method for the treatment of the electronic degrees of freedom, and replace
it with a tight-binding (TB) model \cite{Colombo05}.
We adopt the TB scheme of Xu {\it et al.}\ \cite{Xu92}, which has been
applied successfully to investigate several low-dimensional carbon systems
\cite{Canning97,Yamaguchi07,Cadelano09,Bonelli09}, as it reproduces well
the experimental bulk equilibrium distance $d_{\rm graph} = 1.4224$~\AA\ of
graphite, and the bulk structure and elastic properties of $sp^3$ diamond.
Since all interatomic interactions vanish at a cutoff distance
$r_c=2.6$~\AA, which is shorter than the interlayer spacing of graphite,
3.35~\AA\ \cite{Zacharia04}, we focus on a single-layer model like we did
for the DFT-LDA model above.

\begin{figure}
\begin{center}
\includegraphics[width=0.5\textwidth,angle=0,clip=]{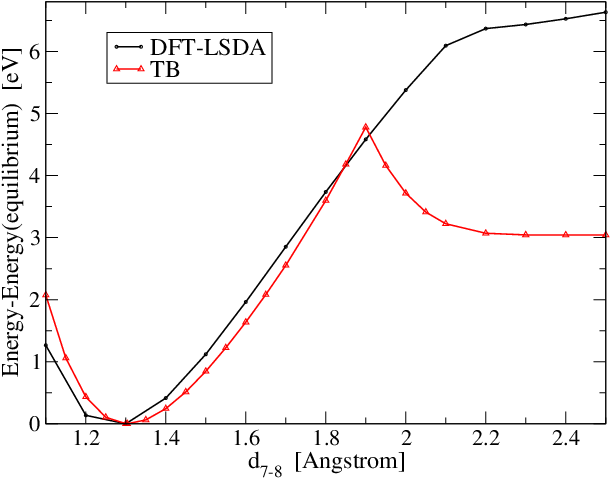}
\end{center}
\caption{\label{compareDFT-TB:fig} (Color online)
  Comparison of the total energy variation for a C$_{14}$ linear chain as a
  function of the (constrained) length $d_{7-8}$ of its central bond
  (between atoms 7 and 8), as obtained with DFT-LSDA and with TB.
}
\end{figure}

To validate the TB force field for spCCs, we compute the total energy of a
C$_{14}$ linear chain as a function of the length $d_{7-8}$ of its central
bond (between atoms 7 and 8), which we keep fixed while all other bonds are
allowed to relax.
Figure~\ref{compareDFT-TB:fig} compares this total energy, referred to its
value at full relaxation, as obtained with DFT-LSDA and with TB.
The two models exhibit significant differences, in particular the TB model
has a level crossing to a dissociative regime above 1.9~\AA, while nothing
of the sort occurs in the DFT-LDA band-structure calculation, which takes
care of level degeneracies by selecting a spin-1 magnetic state.
This problem with the TB model is characteristic of unsaturated conditions,
while, in close-shell electronic configurations, dissociation is more
regular.
In general, the TB model is likely to be more ``fragile'' than a more
realistic DFT-LDA.
In turn the latter is also expected to be less strongly bounded than real
spCCs, due to missing long-range attractive Van-der-Waals polarization
correlation effects.
We must then conclude that all quantitative stability evaluations based on
the TB model are underestimates of the actual stability in experiment.
Periodic boundary conditions (matching the zero-temperature lattice
parameter of graphene) suppress long wavelength fluctuations, resulting in
a stabilization of the thermal fluctuations, which in the thermodynamical
limit would make the 1D - 2D structure unstable.

We run microcanonical (constant-energy) TB molecular dynamics (TBMD)
simulations, since the temperature fluctuations are small enough ($\simeq
10\%$) in a sample of this size for temperature to be considered a fairly
well defined quantity.
The advantage of the microcanonical ensemble is that no thermostat
artifacts, and in particular no dissipative term as in Langevin or
Nos\'e-Hoover thermostats, can affect the atomic dynamics, allowing for a
full account of local fluctuations to produce whatever bond breaking they
may lead to.
A disadvantage of the constant-energy MD is that, if a significant bonding
breakdown of a part of the nanostructure occurs, the corresponding
potential-energy increase occurs at the expense of the kinetic energy, thus
the system may artificially cool down, thus hindering further
decomposition.
In practice, this problem has little importance for the system size
considered.
We use a time step of 0.5~fs, small enough to guarantee a rigorous global
energy conservation within 0.02~eV, or 0.001\%.

We run extended simulations of the model nh-C$_{10}$ structure at different
temperatures, starting with a sampling of initial conditions.
We generate starting configurations by beginning with the fully relaxed
configuration and running three successive equilibration runs (0.1~ps,
0.1~ps and 0.5~ps), with randomized initial velocities, taken from a
Gaussian distribution matching the Boltzmann distribution at the target
temperature.
In the figures we indicate the resulting temperature, with an error bar
appropriate to the determination of the average along the whole simulation,
(thus not estimating the instantaneous temperature fluctuations, which are
much larger).
In the determination of this average temperature, we drop the first 100~fs,
to minimize the systematic oscillations induced by starting with initial
random velocities with little correlation to the forces.

\begin{figure}
\begin{center}
  \subfigure[$t=0$~ps]{\includegraphics[width=0.22\textwidth,angle=0,clip=]{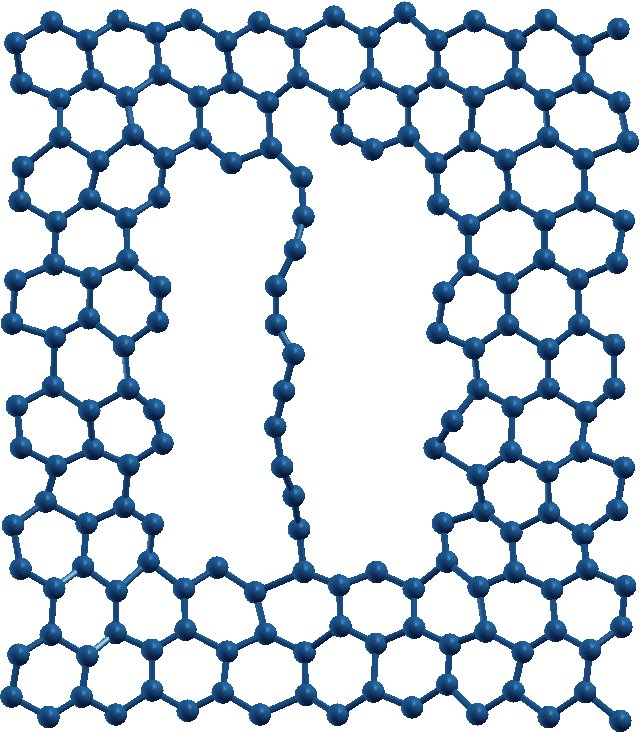}\label{4000_1:fig}}
  \hfill
  \subfigure[$t=1.5$~ps]{\includegraphics[width=0.22\textwidth,angle=0,clip=]{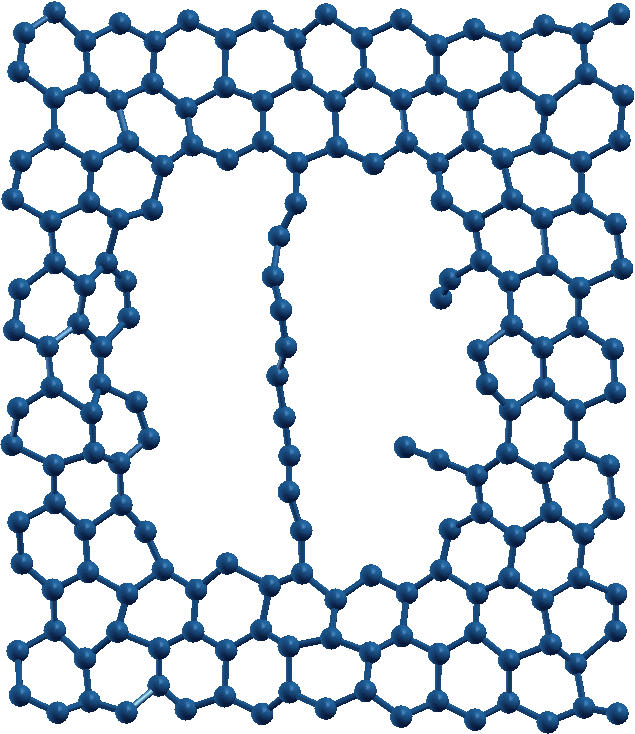}\label{4000_2:fig}}
  \hfill
  \subfigure[$t=2.5$~ps]{\includegraphics[width=0.22\textwidth,angle=0,clip=]{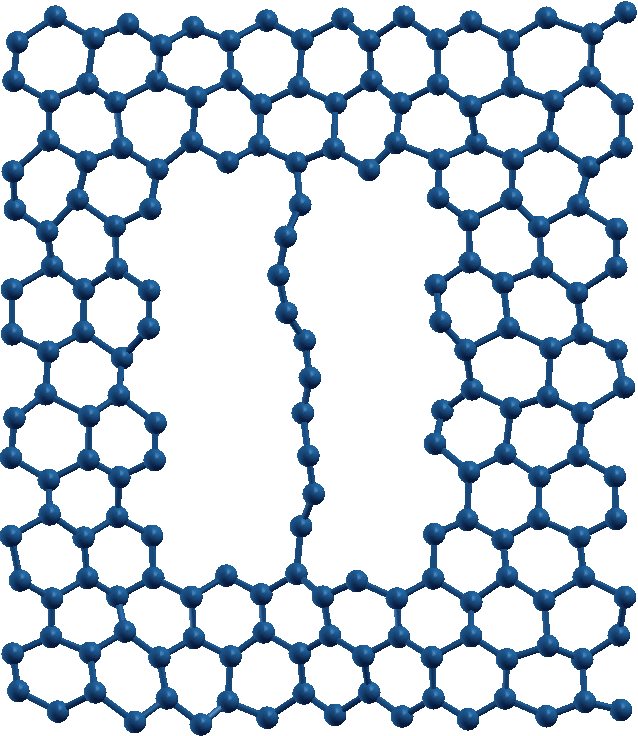}\label{4000_3:fig}}
  \hfill
  \subfigure[$t=2.7$~ps]{\includegraphics[width=0.22\textwidth,angle=0,clip=]{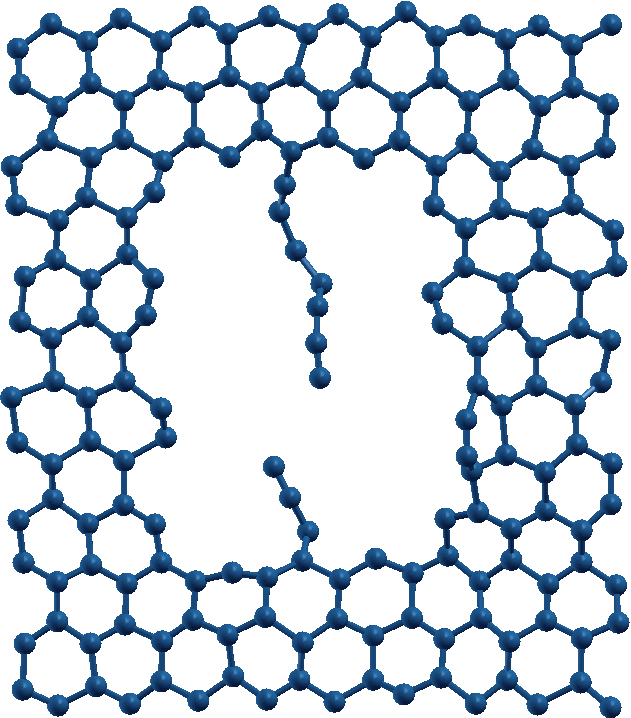}\label{4000_4:fig}}
\end{center}
\caption{\label{dynam_4000K:fig} (Color online)
  Successive snapshots of a sample TBMD simulation at a temperature
  $T=(3991\pm 10)$\,K of nh-C$_{10}$.
}
\end{figure}

\begin{figure}
\begin{center}
 \includegraphics[width=0.5\textwidth,angle=0,clip=]{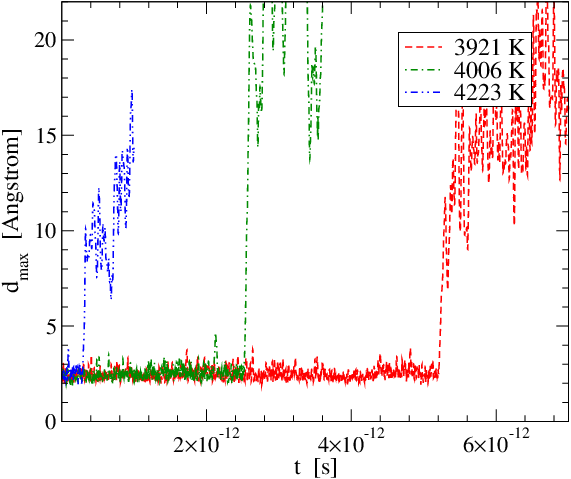}
\end{center}
\caption{\label{max_bond_length:fig} (Color online)
  The time dependence of the longest bond length $d_{\rm max}$,
  Eq.~\eqref{dmax:eq}, of the C$_{10}$ chain in four simulations at
  different temperature.
}
\end{figure}

Figure~\ref{dynam_4000K:fig} displays successive frames of an example
simulation illustrating that edge reconstruction processes --
Fig.~\ref{4000_2:fig} -- often occur before the earliest spCC breakdown
event -- Fig.~\ref{4000_4:fig}.
To monitor these breakdown events, a clear indicator is the longest C-C bond
length relative to the 11 bonds of the C$_{10}$ chain and of the chain ends
to the attached nh edge atoms:
\begin{equation}\label{dmax:eq}
d_{\rm max} =\max_{i=0,1,..10} d_{\rm C_i-C_{i+1}}
\,.
\end{equation}
Figure~\ref{max_bond_length:fig} reports the time dependency of $d_{\rm
  max}$ following 4 independent simulations carried out at different
temperature, with different initial states.
Chain breakdown, as happens between frames~\ref{4000_3:fig} and
\ref{4000_4:fig}, is signaled by the rapid increase of one of the bond
lengths beyond $5$~\AA.
The spCC breakdown may be followed by recombination of the chain into the
nh edge, or even expulsion of a section of the spCC into vacuum.

As suggested by Fig.~\ref{max_bond_length:fig}, spCC breakdown occurs, on
average, earlier and earlier for increasing temperature.
By repeating the numerical simulations for different initial conditions but
similar temperature we estimate an average decay rate by averaging the
inverse times before decay.
The average decomposition rate time $\bar k$ is an increasing function of
temperature.
In simulations done at substantially lower temperature than the ones
considered in Fig.~\ref{max_bond_length:fig}, one would need to wait too
long to observe decomposition, while at much higher temperature
decomposition occurs immediately after start.

\begin{figure}
\begin{center}
  \includegraphics[width=0.5\textwidth,angle=0,clip=]{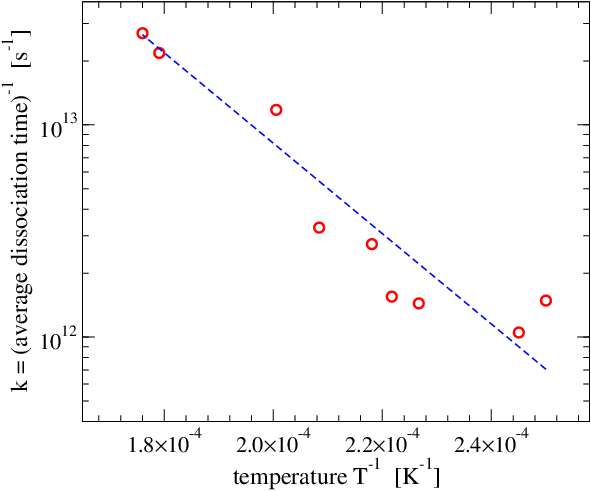}
\end{center}
\caption{\label{arrhenius:fig} (Color online)
  Circles: the average inverse time before the breakdown of the C$_{10}$
  spCC, as a function of inverse temperature.
  Dashed line: a Arrhenius fit, Eq.~\eqref{arrhenius:eq}.
}
\end{figure}

If one can assume that one type of process (bond breaking) dominates over
all decomposition channels, the decomposition rate ${\bar k}$ is expected
to be in the Arrhenius form
\begin{equation}\label{arrhenius:eq}
{\bar k} = A \exp\left(-\frac{E_a}{k_{\rm B}T}\right)
\,.
\end{equation}
We estimate the attempt frequency $A$ and the effective energy barrier
$E_a$ of the TBMD model by fitting the Arrhenius plot in
Fig.~\ref{arrhenius:fig}.
The value $E_a=4.2$~eV matches the TB breakup of
Fig.~\ref{compareDFT-TB:fig}: this effective activation barrier is surely
an underestimation of the actual barrier against breakup.
The estimated attempt rate $A=1.5\times 10^{17}$~s$^{-1}$, although probably
slightly overestimated, reflects the high number of breakup channels
available for decomposition of the spCC \cite{supplementary:note}.
If we assume the computed values for $E_a$ and $A$, we extrapolate thermal
decay rates of spCCs in $sp^2$ carbon of the order of ${\bar k} = 7.5\times
10^{-5}$~s$^{-1}$ at 1000~K, and ${\bar k} = 1.5\times 10^{-54}$~s$^{-1}$
at room-temperature (300~K).
Overall, the calculations of the present section confirm a substantial
stability of spCCs in the solid state and in vacuum.
In the lab, whenever $sp-sp^2$ carbon is not kept in vacuum, chemical decay
mechanisms are therefore likely to overcome the thermal ones.

\section{Discussion and conclusion}\label{conclusion:sec}

The present work collects extensive investigation of the geometry,
electronic structure, magnetic properties, and dynamical stability of
spCCs attached to $sp^2$ graphitic fragments.
When a spCC binds to zig-zag graphene edges, its polyynic character is
attenuated to a value intermediate between those typical of cumulenes and
polyynes.
The attachment of a spCC to the graphene edge is very stable: we predict
stabilization energies near $6$~eV per bond between each chain end and
the $sp^2$ regions.
Thermal excitations typically break bonds along the spCC with similar
probability to those formed with the graphene edge, which indicates a very
solid attachment.

Odd spCCs in the nh display a metallic behavior, with at least one band
pinned to the Fermi energy, while even spCCs have little overlap with the
states at the Fermi level \cite{Standley08,YLi08,Avouris07,Zhang11}.
The partly filled states of odd-$n$ spCCs are associated to
nonzero magnetization related to a spin triplet state of the $\pi$ bonds.
Even-$n$ spCCs are instead insulating and non-magnetic, and in this context
only the magnetic moments of the graphene edge contribute to the magnetism
of the nh-C$_{2m}$ structures, and only the bulk graphene states provide
conducting bands.

We compute also the vibrational modes, and specifically the optical
C$\equiv$C stretching modes which emerge as a characteristic signature of
spCCs in Raman and IR spectroscopies.
Our calculations show that the vibrational frequencies can be quite
substantially red-shifted when spCCs are kept under tensile stress.
%
%
Indeed, the weaker stability and correspondingly faster decomposition rate
of strained spCCs is likely to play a significant role in the overall blue
shift of the $sp$-carbon peak in the Raman spectrum of the decaying spCCs
in cluster-assembled $sp$-$sp^2$ film \cite{Ravagnan09}.

\section*{Acknowledgments}
We are grateful to L. Ravagnan, P. Milani, E. Cinquanta, and Z. Zanolli
for invaluable discussions.
The research leading to these results has received funding from the
European Community's Seventh Framework Programme (FP7/2007-2013) under
grant agreement No.~211956 (ETSF-i3).
We acknowledge generous supercomputing support from CILEA.

\appendix
\section{Magnetism at the edge of the empty nh}\label{mag_nh_edge:sec}

\begin{figure}
\centerline{
\includegraphics[width=0.55\textwidth,angle=0,clip=]{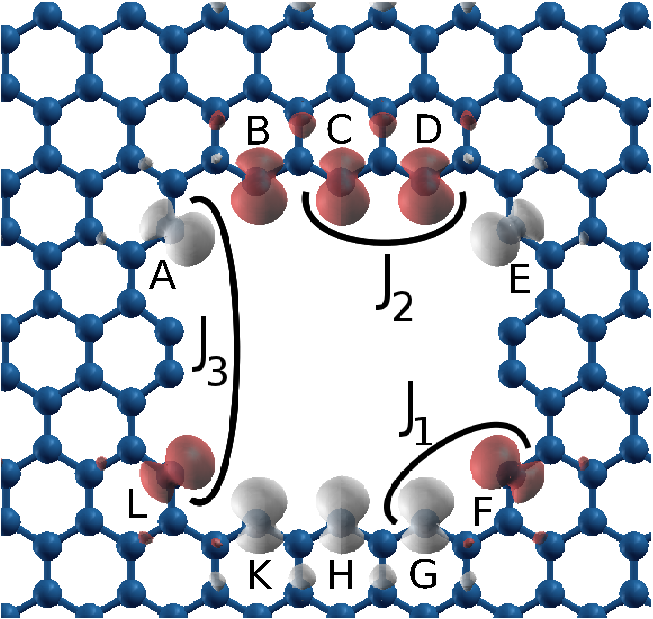}
}
\caption{\label{j_hole:fig} (Color online)
  Magnetization-density isosurfaces at $+0.01\,\mu_B/a_0^3$ (dark/red) and
  $-0.01\,\mu_B/a_0^3$ (clear gray) for the magnetic ground state of the
  unpaired-spin electrons localized at the zig-zag edges of the empty nh.
  Letters label the spin-carrying atomic sites.
  Arcs mark all symmetry-independent nearest-neighbor Ising-type magnetic
  couplings $J_i$, see Eq.~(\ref{ising_model}).
}
\end{figure}

As magnetism is intrinsic of the zig-zag graphene edge, thus even a empty
nanohole exhibits a range of magnetic states similar to those arising in
the presence of even-$n$ spCCs.
The zig-zag edge atoms involved in magnetism are clearly identified in
Fig.~\ref{j_hole:fig}.
Also here, each atomic site A--L carries a magnetization close to
$1\,\mu_{\rm B}$.
As illustrated in Fig.~\ref{j_hole:fig}, even though the number of
spin-carrying atoms is larger than in the case of Sect.~\ref{mag_c6nh:sec},
only three independent nearest-neighbor interactions need to be considered.
Two of them, $J_1$ and $J_3$, represent interaction between two edges
belonging to different sublattices (where we expect an anti-ferromagnetic
character).
In contrast, the coupling $J_2$ accounts for the interaction of spins
within the same edge, and is therefore expected to be ferromagnetic,
according to Ref.~\cite{Yu08}.

\begin{figure}
\begin{center}
  \subfigure[
  {\it Ground State}\newline$E_{\rm tot}=E_{\rm gs}=-13017.539$~eV\newline$M_{\rm tot}=0.00$\newline$M_{\rm abs}=11.76$]
  {\includegraphics[width=0.30\textwidth,angle=0,clip=]{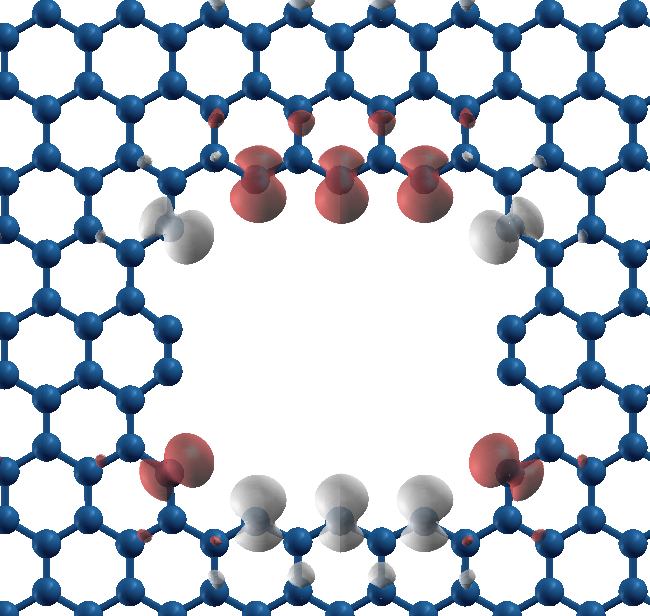}\label{hole_mag1:fig}}
  \hspace{0.03\textwidth}
  \subfigure[$E_{\rm tot}=E_{\rm gs}+52$~meV\newline$M_{\rm tot}=0.00$\newline$M_{\rm abs}=10.05$]{\includegraphics[width=0.30\textwidth,angle=0,clip=]{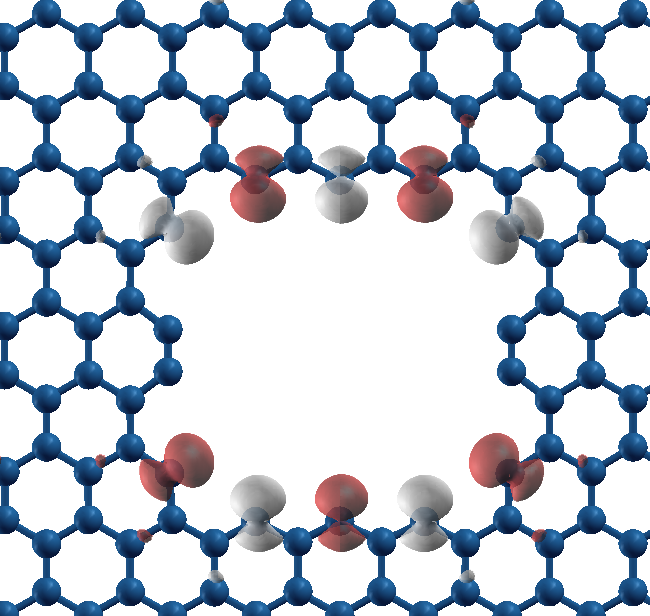}\label{hole_mag4:fig}}
  \hspace{0.03\textwidth}
  \subfigure[$E_{\rm tot}=E_{\rm gs}+72$~meV\newline$M_{\rm tot}=-2.00$\newline$M_{\rm abs}=9.71$]{\includegraphics[width=0.30\textwidth,angle=0,clip=]{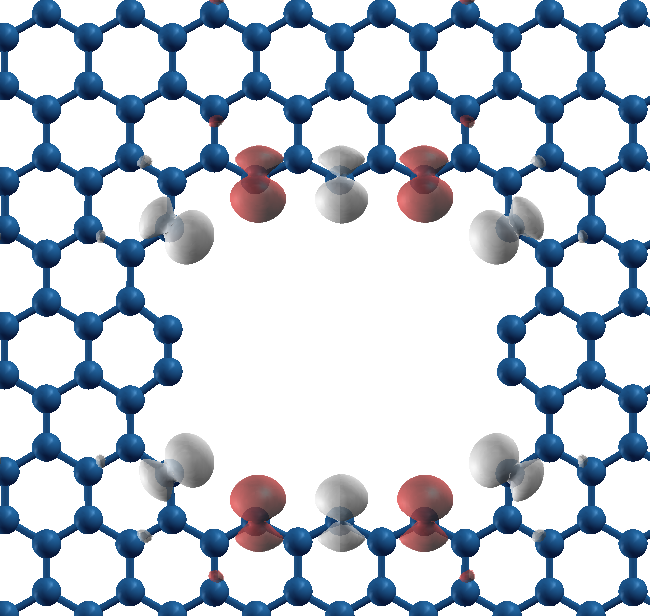}\label{hole_mag3:fig}}
  \hspace{0.03\textwidth}
  \subfigure[$E_{\rm tot}=E_{\rm gs}+86$~meV\newline$M_{\rm tot}=2.00$\newline$M_{\rm abs}=10.77$]{\includegraphics[width=0.30\textwidth,angle=0,clip=]{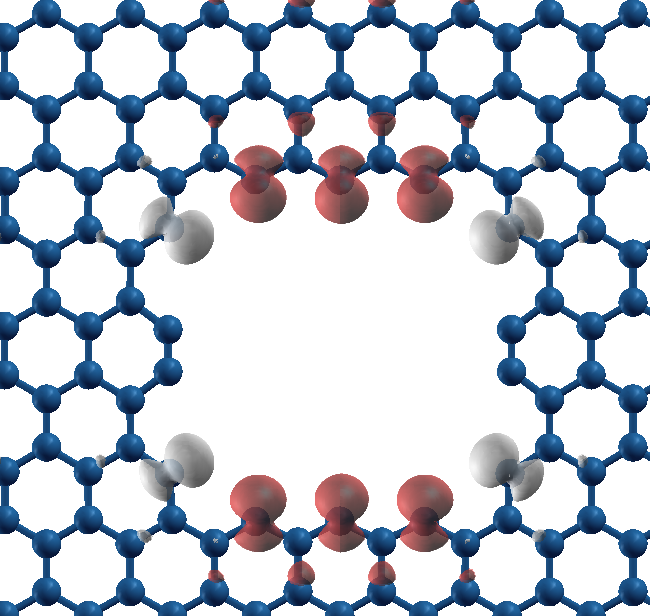}\label{hole_mag2:fig}}
  \hspace{0.03\textwidth}
  \subfigure[$E_{\rm tot}=E_{\rm gs}+513$~meV\newline$M_{\rm tot}=0.00$\newline$M_{\rm abs}=11.52$]{\includegraphics[width=0.30\textwidth,angle=0,clip=]{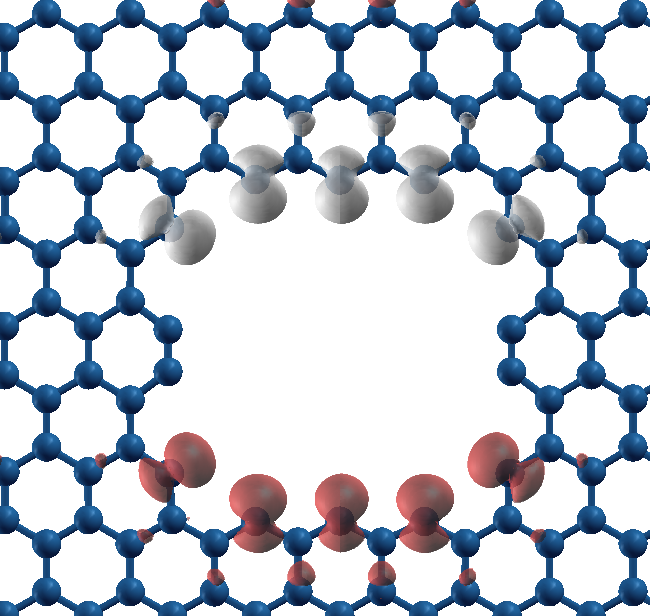}\label{hole_mag5:fig}}
  \hspace{0.03\textwidth}
  \subfigure[$E_{\rm tot}=E_{\rm gs}+565$~meV\newline$M_{\rm tot}=10.00$\newline$M_{\rm abs}=10.77$]{\includegraphics[width=0.30\textwidth,angle=0,clip=]{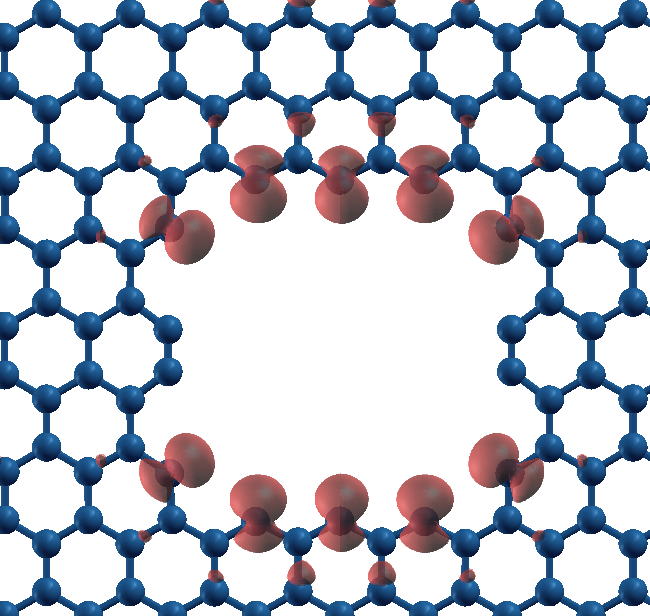}\label{hole_mag6:fig}}
\end{center}
\caption{\label{holemag:fig} (Color online)
  Magnetic positive (dark/red) $+0.01\,\mu_{\rm B}/a_0^3$
  and negative (clear/gray)    $-0.01\,\mu_{\rm B}/a_0^3$
  isosurfaces for the empty nanohole structures.
}
\end{figure}

\begin{table}
\begin{center}
\begin {tabular}{c|c|c}
\hline \hline
Ising Parameter & Value [meV] & Standard deviation [meV]\\ \hline
$E_0-E_{\rm gs}$ 	& $300$ & $12$ \\
$J_1$ 			& $-248$ & $12$ \\
$J_2$			& $10$ & $12$ \\
$J_3$ 			& $-53$ & $19$\\
\hline \hline
\end{tabular}
\end{center}
\caption{\label{j_hole:tab} (Color online)
  The individual Ising-model parameters computed for the nh magnetic
  configurations.
}
\end{table}

As done in Sect.~\ref{mag_c6nh:sec} for the nh-C$_6$ structures, we make a
linear fit of all considered magnetic configurations (shown in
Fig.~\ref{holemag:fig}), to evaluate the values of the $J_i$ and of $E_0$.
In the Ising model, the total energy is written as:
\begin{eqnarray}\label{ising_hole}
E_{\rm tot} = E_0 + E_{\rm spin} = E_0&-&
J_1\left(S_A S_B+S_D S_E+S_F S_G+S_K S_L\right) + \nonumber\\
&-&J_2\left(S_B S_C+S_C S_D+S_G S_H+S_H S_K\right) + \nonumber\\
&-&J_3\left(S_A S_L + S_E S_F\right)
\,.
\end{eqnarray}
The result of the linear fit of the energies of the magnetic configurations
of Fig.~\ref{holemag:fig} is reported in Table~\ref{j_hole:tab}.
Like in nh-C$_6$, $J_1$ is much larger than the other antiferromagnetic
coupling $J_3$.
Indeed, the $J_1$ interaction is very similar to the one obtained in the
calculations with the nh-C$_6$ structure, see Table~\ref{j_carb6:tab}.
Moreover, like in the nh-C$_6$ case, given the relevant error bar the
weakly ferromagnetic $J_2$ is in fact compatible with a null value, which
is somewhat surprising for neighboring atoms along the same zig-zag edge.

\begin{figure}
\centerline{
\includegraphics[width=0.75\textwidth,angle=0,clip=]{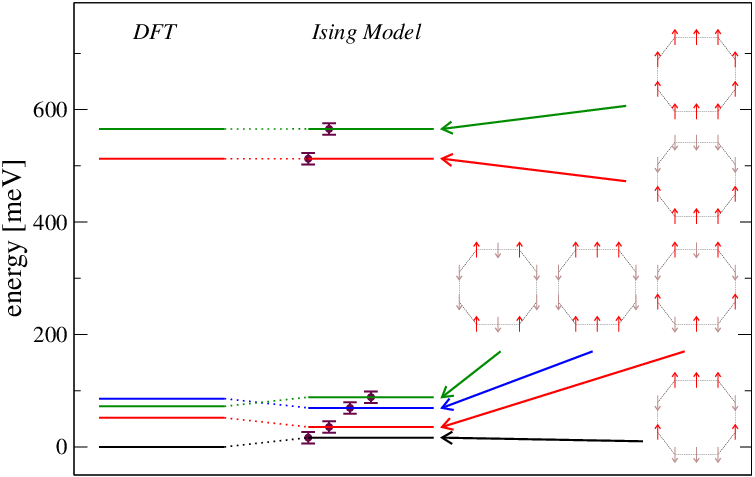}
}
\caption{\label{comp_hole:fig} (Color online)
  A comparison of the DFT-LSDA magnetic energy levels of the structures of
  Fig.~\ref{holemag:fig}, with those obtained based on the Ising model,
  Eq.~\eqref{ising_hole}, fitted on the DFT-LSDA energies for the empty nh.
}
\end{figure}

Figure~\ref{comp_hole:fig} compares the DFT-LSDA energy levels and those
obtained using the fitted Ising model.
Here the fit agrees better than in the nh-C$_6$ case, but clearly the role
of the $J_2$ coupling is contradictory, which explains its small value.

\section*{References}


\begin{thebibliography}{10}

\bibitem{Cataldo05}
{ {\it Polyynes: Synthesis, Properties, and Applications}, edited by F.\
  Cataldo (CRC, Taylor\&Francis, London, 2005)}.

\bibitem{polyynes:note}
{ Since ``polyyne'' indicates an alternating single-triple-bond carbon chain,
  we use the more generic ``$sp$ carbon chain'' to include also undimerized
  species, such as cumulenes or odd-$n$ C$_n$ chains}.

\bibitem{ElGoresy68}
{ A.\ El Goresy and G.\ Donnay, Science {\bf 161}, 363 (1968)}.

\bibitem{Kroto92}
{ H.\ Kroto, Carbon {\bf 30}, 1139 (1992)}.

\bibitem{Duley09}
{ W.\ W.\ Duley, and A.\ Hu, Astrophys.\ J.\ {\bf 698}, 808 (2009)}.

\bibitem{Bundy96}
{ F.\ P.\ Bundy, W.\ A.\ Bassett, M.\ S.\ Weathers, R.\ J.\ Hemley, H.\ K.\
  Mao, and A.\ F.\ Goncharov, Carbon {\bf 34}, 141 (1996)}.

\bibitem{Baughman06}
{ R.\ H.\ Baughman, Science {\bf 312}, 1009 (2006)}.

\bibitem{Matsuda84}
{ H.\ Matsuda, H.\ Nakanishi, and M.\ Kato, J.\ Polym.\ Lett.\ Ed.\ {\bf 22},
  107 (1984)}.

\bibitem{Kudryavtsev69}
{ Y.\ P.\ Kudryavtsev, {\it Progress of Polymer Chemistry}, 87, Nauka (Moscow,
  1969)}.

\bibitem{Cataldo99}
{ F.\ Cataldo, D.\ Capitani, Materials Chem.\ Phys.\ {\bf 59}, 225 (1999)}.

\bibitem{Mohr03}
{ W.\ Mohr, J.\ Stahl, F.\ Hampel, and J.\ A.\ Gladysz, Chem.\ Eur.\ J.\ {\bf
  9}, 3324 (2003)}.

\bibitem{Zhao03}
{ X.\ Zhao, Y.\ Ando, Y.\ Liu, M.\ Jinno, and T.\ Suzuki, Phys.\ Rev.\ Lett.\
  {\bf 90}, 187401 (2003)}.

\bibitem{Liu03}
{ Y.\ Liu {\it et al.}, Phys.\ Rev.\ B {\bf 68}, 125413 (2003)}.

\bibitem{Inoue10}
{ K.\ Inoue, R.\ Matsutani, T.\ Sanada, and K.\ Kojima, Carbon {\bf 48}, 4209
  (2010)}.

\bibitem{Rice10}
{ C.\ A.\ Rice, V.\ Rudnev, R.\ Dietsche, and J.\ P.\ Maier, Astron.\ J.\ {\bf
  140}, 203 (2010)}.

\bibitem{Kijima96}
{ M.\ Kijima, T.\ Toyabe, and H.\ Shirakawa, Chem.\ Commun.\ {\bf 19}, 2273
  (1996)}.

\bibitem{Heimann99}
{ R.\ B.\ Heimann, S.\ E.\ Evsyukov, and L.\ Kavan, {\it Carbyne and Carbynoid
  Structures} (Kluwer, Dordrecht, 1999)}.

\bibitem{Tsuji03}
{ M.\ Tsuji, S.\ Kuboyama, T.\ Matsuzaki, and T.\ Tsuji, Carbon {\bf 41}, 2141
  (2003)}.

\bibitem{Cataldo04}
{ F.\ Cataldo, Tetrahedron Lett.\ {\bf 45}, 141 (2004)}.

\bibitem{Yamada91}
{ K.\ Yamada, H.\ Kunishige, and A.\ B.\ Sawaoka, Naturwiss {\bf 78}, 450
  (1991)}.

\bibitem{Ohmura97}
{ K.\ Ohmura, M.\ Kijima, and H.\ Shirakawa, Synth.\ Metals {\bf 84}, 417
  (1997)}.

\bibitem{Kijima97}
{ M.\ Kijima, Recent Res.\ Devel.\ Pure Appl.\ Chem.\ {\bf 1}, 27 (1997)}.

\bibitem{Ravagnan02}
{ L.\ Ravagnan, F.\ Siviero, C.\ Lenardi, P.\ Piseri, E.\ Barborini, and P.\
  Milani, Phys.\ Rev.\ Lett.\ {\bf 89}, 285506 (2002)}.

\bibitem{Ravagnan07}
{ L.\ Ravagnan, P.\ Piseri, M.\ Bruzzi, S.\ Miglio, G.\ Bongiorno, A.\ Baserga,
  C.\ S.\ Casari, A.\ Li Bassi, C.\ Lenardi, Y.\ Yamaguchi, T.\ Wakabayashi,
  C.\ E.\ Bottani, and P.\ Milani, Phys.\ Rev.\ Lett.\ {\bf 98}, 216103
  (2007)}.

\bibitem{Ravagnan09}
{ L.\ Ravagnan, N.\ Manini, E.\ Cinquanta, G.\ Onida, D.\ Sangalli, C.\ Motta,
  M.\ Devetta, A.\ Bordoni, P.\ Piseri, and P.\ Milani, Phys.\ Rev.\ Lett.\
  {\bf 102}, 245502 (2009)}.

\bibitem{Troiani03}
{ H.\ E.\ Troiani, M.\ Miki-Yoshida, G.\ A.\ Camacho-Bragado, M.\ A.\ L.\
  Marques, A.\ Rubio, J.\ A.\ Ascencio, and M.\ Jose-Yacaman, Nano Lett.\ {\bf
  3}, 751 (2003)}.

\bibitem{Jin09}
{ C.\ Jin, H.\ Lan, L.\ Peng, K.\ Suenaga, and S.\ Iijima, Phys.\ Rev.\ Lett.\
  {\bf 102}, 205501 (2009)}.

\bibitem{Chuvilin09}
{ A.\ Chuvilin, J.\ C.\ Meyer, G.\ Algara-Siller, and U.\ Kaiser, New J.\
  Phys.\ {\bf 11}, 083019 (2009)}.

\bibitem{Mikhailovskij09}
{ I.\ M.\ Mikhailovskij, E.\ V.\ Sadanov, T.\ I.\ Mazilova, V.\ A.\
  Ksenofontov, and O.\ A.\ Velicodnaja, Phys.\ Rev.\ B {\bf 80}, 165404
  (2009)}.

\bibitem{Zeng10}
{ M.\ G.\ Zeng, L.\ Shen, Y.\ Q.\ Cai, Z.\ D.\ Sha, and Y.\ P.\ Feng, Appl.\
  Phys.\ Lett.\ {\bf 96}, 042104 (2010)}.

\bibitem{Chalifoux09}
{ W.\ A.\ Chalifoux and R.\ R.\ Tykwinski, C.\ R.\ Chimie {\bf 12}, 341
  (2009)}.

\bibitem{Hobi10}
{ E.\ Hobi Jr., R.\ B.\ Pontes, A.\ Fazzio, and A.\ J.\ R. da Silva, Phys.\
  Rev.\ B {\bf 81}, 201406 (2010)}.

\bibitem{Akdim11}
{ B.\ Akdim and R.\ Pachter, ACSNano {\bf 5}, 1769 (2011)}.

\bibitem{Hu11}
{ Y.\ H.\ Hu, J.\ Phys.\ Chem.\ C {\bf 115}, 1843 (2011)}.

\bibitem{Ravagnan11}
{ L.\ Ravagnan, T.\ Mazza; G.\ Bongiorno, M.\ Devetta, M.\ Amati, P.\ Milani,
  P.\ Piseri, M.\ Coreno, C.\ Lenardi, F.\ Evangelista and P.\ Rudolf, Chem.\
  Commun.\ {\bf 47}, 2952 (2011)}.

\bibitem{Erdogan11}
{ E.\ Erdogan1, I.\ Popov, C.\ G.\ Rocha, G.\ Cuniberti, S.\ Roche, and G.\
  Seifert, Phys.\ Rev.\ B {\bf 83}, 041401 (2011)}.

\bibitem{Makarova01}
{ T.\ L.\ Makarova, B.\ Sundqvist, R.\ H\"ohne, P.\ Esquinazi, Y.\ Kopelevich,
  P.\ Scharff, V.\ A.\ Davydov, L.\ S.\ Kashevarova, and A.\ V.\ Rakhmanina,
  Nature (London) {\bf 413}, 716 (2001)}.

\bibitem{Esquinazi02}
{ P.\ Esquinazi, A.\ Setzer, R.\ Höhne, C.\ Semmelhack, Y.\ Kopelevich, D.\
  Spemann, T.\ Butz, B.\ Kohlstrunk, and M.\ L\"osche, Phys.\ Rev.\ B {\bf 66},
  024429 (2002)}.

\bibitem{Coey02}
{ J.\ M.\ D.\ Coey, M.\ Venkatesan, C.\ B.\ Fitzgerald, A.\ P.\ Douvalis, and
  I.\ S.\ Sanders, Nature {\bf 420}, 156 (2002)}.

\bibitem{Esquinazi03}
{ P.\ Esquinazi, D.\ Spemann, R.\ H\"ohne, A.\ Setzer, K.-H.\ Han, and T.\
  Butz, Phys.\ Rev.\ Lett.\ {\bf 91}, 227201 (2003)}.

\bibitem{Ohldag07}
{ H.\ Ohldag, T.\ Tyliszczak, R.\ H\"ohne, D.\ Spemann, P.\ Esquinazi, M.\
  Ungureanu, and T.\ Butz, Phys.\ Rev.\ Lett.\ {\bf 98}, 187204 (2007)}.

\bibitem{Klein99}
{ D.\ J.\ Klein and L.\ Bytautas, J.\ Phys.\ Chem A {\bf 103}, 5196 (1999)}.

\bibitem{Son06b}
{ Y.\ W.\ Son, M.\ Cohen and S.\ Louie, Nature {\bf 444}, 347 (2006)}.

\bibitem{Yazyev08}
{ O.\ V.\ Yazyev, and M.\ I.\ Katsnelson, Phys.\ Rev.\ Lett.\ {\bf 100}, 047209
  (2008)}.

\bibitem{Uchoa08}
{ B.\ Uchoa, V.\ N.\ Kotov, N.\ M.\ R.\ Peres, and A.\ H.\ Castro Neto, Phys.\
  Rev.\ Lett.\ {\bf 101}, 026805 (2008)}.

\bibitem{Pisani08}
{ L.\ Pisani, B.\ Montanari, and N.\ M.\ Harrison, New J.\ Phys.\ {\bf 10},
  033002 (2008)}.

\bibitem{Zanolli10}
{ Z.\ Zanolli and J.-C.\ Charlier, Phys.\ Rev.\ B {\bf 81}, 165406 (2010)}.

\bibitem{Son06}
{ Y.\ W.\ Son, M.\ L.\ Cohen, and S.\ G.\ Louie, Phys.\ Rev.\ Lett.\ {\bf 97},
  216803 (2006); Nature {\bf 444}, 347 (2006)}.

\bibitem{Yang08b}
{ L.\ Yang, M.\ L.\ Cohen, and S.\ G.\ Louie, Phys.\ Rev.\ Lett.\ {\bf 101},
  186401 (2008)}.

\bibitem{Standley08}
{ B.\ Standley, W.\ Bao, H.\ Zhang, J.\ Bruck, C.\ N.\ Lau, and M.\ Bockrath,
  Nano Lett.\ {\bf 8}, 3345 (2008)}.

\bibitem{YLi08}
{ Y.\ Li, A.\ Sinitskii, and J.\ M.\ Tour, Nature Mat.\ {\bf 7}, 966 (2008)}.

\bibitem{DasSarma01}
{ S.\ Das Sarma, Am.\ Scientist {\bf 89}, 516 (2001)}.

\bibitem{Casari04}
{ C.\ S.\ Casari, A.\ Li Bassi, L.\ Ravagnan, F.\ Siviero, C.\ Lenardi, P.\
  Piseri, G.\ Bongiorno, C.\ E.\ Bottani, and P.\ Milani, Phys.\ Rev.\ B {\bf
  69}, 075422 (2004)}.

\bibitem{Cataldo10}
{ F.\ Cataldo, L.\ Ravagnan, E.\ Cinquanta, I.\ E.\ Castelli, N.\ Manini, G.\
  Onida, and P.\ Milani, J.\ Phys.\ Chem.\ B {\bf 114}, 14834 (2010)}.

\bibitem{Cinquanta11}
{ E.\ Cinquanta, L.\ Ravagnan, I.\ E.\ Castelli, F.\ Cataldo, N.\ Manini, G.\
  Onida, and P.\ Milani, submitted to J.\ Chem.\ Phys}.

\bibitem{Pickett89}
{ W.\ E.\ Pickett, Comput.\ Phys.\ Rep.\ {\bf 9}, 115 (1989)}.

\bibitem{B3LYP}
{ A.\ D.\ Becke, J.\ Chem.\ Phys.\ {\bf 98}, 5648 (1993)}.

\bibitem{PBE96}
{ J.\ P.\ Perdew, K.\ Burke, and M.\ Ernzerhof, Phys.\ Rev.\ Lett.\ {\bf 77},
  3865 (1996)}.

\bibitem{Xu04}
{ X.\ Xu and W.\ A.\ Goddard III, J.\ Chem.\ Phys.\ {\bf 121}, 4068 (2004)}.

\bibitem{espresso2009}
{ P.\ Giannozzi, S.\ Baroni, N.\ Bonini, M.\ Calandra, R.\ Car, C.\ Cavazzoni,
  D.\ Ceresoli, G.\ L.\ Chiarotti, M.\ Cococcioni, I.\ Dabo, A.\ Dal Corso, S.
  de Gironcoli, S.\ Fabris, G.\ Fratesi, R.\ Gebauer, U.\ Gerstmann, C.\
  Gougoussis, A.\ Kokalj, M.\ Lazzeri, L.\ Martin-Samos, N.\ Marzari, F.\
  Mauri, R.\ Mazzarello, S.\ Paolini, A.\ Pasquarello, L.\ Paulatto, C.\
  Sbraccia, S.\ Scandolo, G.\ Sclauzero, A.\ P.\ Seitsonen, A.\ Smogunov, P.\
  Umari, and R.\ M.\ Wentzcovitch, J.\ Phys.: Condens.\ Matter {\bf 21}, 395502
  (2009)}.

\bibitem{Vanderbilt90}
{ D.\ Vanderbilt, Phys.\ Rev.\ B {\bf 41}, 7892 (1990)}.

\bibitem{Favot99}
{ F.\ Favot and A.\ Dal Corso, Phys.\ Rev.\ B {\bf 60}, 11427 (1999)}.

\bibitem{BLAdefinition:note}
{ The BLA measures the degree of dimerization and, excluding the terminal
  bonds, can be defined as $ \frac 12 \left[ \sum_{j=1}^{n_o} (d_{2j-1}+
  d_{n-(2j-1)})/n_o -\sum_{j=1}^{n_e} (d_{2j} +d_{n-2j})/n_e \right] $, with
  $d_i=|\vec r_i-\vec r_{i+1}|$, $n_o=(n+2)/4$, and $n_e=n/4$ (taken as integer
  part of these fractions)}.

\bibitem{Castelli11}
{ I.\ E.\ Castelli and N.\ Manini, arXiv:1106.0689}.

\bibitem{Cahangirov10}
{ S.\ Cahangirov, M.\ Topsakal, and S.\ Ciraci, Phys.\ Rev.\ B {\bf 82}, 195444
  (2010)}.

\bibitem{Okada08}
{ S.\ Okada, Phys.\ Rev.\ B {\bf 77}, 041408 (2008)}.

\bibitem{Yu08}
{ D.\ Yu, E.\ M.\ Lupton, M.\ Liu, W.\ Liu, and F.\ Liu, Nano Res.\ {\bf 1}, 56
  (2008)}.

\bibitem{Fujita96}
{ M.\ Fujita, K.\ Wakabayashi, K.\ Nakada, and K.\ Kusakabe, J.\ Phys.\ Soc.\
  Jpn.\ {\bf 65}, 1920 (1996)}.

\bibitem{Kusakabe03}
{ K.\ Kusakabe and M.\ Maruyama, Phys.\ Rev.\ B {\bf 67}, 092406 (2003)}.

\bibitem{Liu09}
{ W.\ Liu, Z.\ F.\ Wang, Q.\ W.\ Shi, J.\ Yang, and F.\ Liu, Phys.\ Rev.\ B
  {\bf 80}, 233405 (2009)}.

\bibitem{Zanolli11}
{ Z.\ Zanolli, G.\ Onida, and J.-C.\ Charlier, ACS Nano {\bf 4}, 5174 (2010)}.

\bibitem{Baroni02}
{ S.\ Baroni, S. de Gironcoli, A.\ Dal Corso, and P.\ Giannozzi, Rev.\ Mod.\
  Phys.\ {\bf 73}, 515 (2001)}.

\bibitem{Tabata06}
{ H.\ Tabata, M.\ Fujii, S.\ Hayashi, T.\ Doi, and T.\ Wakabayashi, Carbon {\bf
  44}, 3168 (2006)}.

\bibitem{Castelli10}
{ I.\ E.\ Castelli, {\it Structural and Magnetic Properties of $sp$-Hybridized
  Carbon}, diploma thesis, (University Milan, 2010),
  \url{http://www.mi.infm.it/manini/theses/castelliMag.pdf}}.

\bibitem{Innocenti10}
{ F.\ Innocenti, A.\ Milani, and C.\ Castiglioni, J.\ Raman Spectrosc.\ {\bf
  41}, 226 (2010)}.

\bibitem{Colombo05}
{ L.\ Colombo, Rivista Nuovo Cimento {\bf 28}, 1 (2005)}.

\bibitem{Xu92}
{ C.\ H.\ Xu, C.\ Z.\ Wang, C.\ T.\ Chan, and K.\ M.\ Ho, J.\ Phys.: Condens.\
  Matter \textbf{4}, 6047 (1992)}.

\bibitem{Canning97}
{ A.\ Canning, G.\ Galli, and J.\ Kim, Phys.\ Rev.\ Lett.\ {\bf 78}, 4442
  (1997)}.

\bibitem{Yamaguchi07}
{ Y.\ Yamaguchi, L.\ Colombo, P.\ Piseri, L.\ Ravagnan, and P.\ Milani, Phys.\
  Rev.\ B {\bf 76}, 134119 (2007)}.

\bibitem{Cadelano09}
{ E.\ Cadelano, P.\ L.\ Palla, S.\ Giordano, and L.\ Colombo, Phys.\ Rev.\
  Lett.\ {\bf 102}, 235502 (2009)}.

\bibitem{Bonelli09}
{ F.\ Bonelli, N.\ Manini, E.\ Cadelano, and L.\ Colombo, Eur.\ Phys.\ J.\ B
  {\bf 70}, 449 (2009)}.

\bibitem{Zacharia04}
{ R.\ Zacharia, H.\ Ulbricht, and T.\ Hertel, Phys.\ Rev.\ B {\bf 69}, 155406
  (2004)}.

\bibitem{supplementary:note}
{ See Supplementary Material, Document No.\ ??, for a few movies illustrating
  several typical decomposition mechanisms}.

\bibitem{Avouris07}
{ Ph.\ Avouris, Z.\ Chen. and V.\ Perebeinos, Nature Nanotec.\ {\bf 2}, 605
  (2007)}.

\bibitem{Zhang11}
{ G.\ P.\ Zhang, X.\ W.\ Fang, Y.\ X.\ Yao, C.\ Z.\ Wang, Z.\ J.\ Ding, and K.\
  M.\ Ho, J.\ Phys.: Condens.\ Matter {\bf 23}, 025302 (2011)}.

\end{thebibliography}

\end{document}